# Targeting the Major Groove of the Palindromic d(GGCGCC)$_2$ Sequence by Oligopeptide Derivatives of Anthraquinone Intercalators

Krystel El Hage [1]*, Giovanni Ribaudo [2]*, Louis Lagardère [3], Alberto Ongaro [2], Philippe H. Kahn [4], Luc Demange [5], Jean-Philip Piquemal [3, 6], Giuseppe Zagotto [7], Nohad Gresh [3]*

1 SABNP, Univ Evry, INSERM U1204, Universite Paris-Saclay, Evry, France
2 Dipartimento di Medicina Molecolare eTraslazionale, Universita degli Studi di Brescia, 25123 Brescia, Italy;
3 LCT, UMR7616 CNRS, Sorbonne Université Paris 75005 France
4 « Retired »
5 Université de Paris, CiTCoM, UMR 8038 CNRS, F-75006, Paris, France
6 The University of Texas at Austin, Department of Biomedical Engineering, Austin, Texas 78705, United States
7 Department of Pharmaceutical and Pharmacological Sciences, University of Padova, Via Marzole , 35131 Padova, Italy.

*e-mail: krystel.elhage@univ-evry.fr; giovanni.ribaudo@unibs.it; nohad.gresh@lct.jussieu.fr

**Abstract.** GC-rich sequences are recurring motifs in oncogenes and retroviruses, and could be targeted by non-covalent major-groove therapeutic ligands. We considered the palindromic sequence d($G_1G_2C_3G_4C_5C_6$)$_2$, and designed several oligopeptide derivatives of the anti-cancer intercalator mitoxantrone. The stability of their complexes with a 18-mer oligonucleotide encompassing this sequence in its center was validated using polarizable molecular dynamics. We report the most salient structural features of two novel compounds, having a dialkylammonium group as a side-chain on both arms. The anthraquinone ring is intercalated in the central d(CpG)$_2$ sequence with its long axis perpendicular to that of the two base-pairs. On each strand, this enables each ammonium group to bind in-register to $O_6$/$N_7$ of the two facing G bases upstream. We subsequently designed tris-intercalating derivatives, each dialkylammonium substituted with a connector to an $N_9$-aminoacridine intercalator extending our target range from six- to a ten-base pair palindromic sequence, d($C_1G_2G_3G_4C_5G_6C_7C_8C_9G_{10}$)$_2$. The structural features of the complex of the most promising derivative are reported. The present design strategy paves the way for designing intercalator-oligopeptide derivatives with an even higher selectivity, targeting an increased number of DNA bases, going beyond ten.

**Introduction.**

The high occurrence of GC-rich sequences in oncogenes was recognized early on (1). They constitute privileged targets for chemotherapeutic agents which could prevent, or interfere with, the early stages of DNA transcription. This was demonstrated for several alkylating anti-cancer drugs, binding covalently to the $N_7$ atom of guanine with an intensity increasing with the number of successive guanines (1). Owing to the toxicity of such drugs and the permanent damage they could inflict on non-oncogenic binding sites, the alternative search for non-covalent, sequence-selective, groove-binding ligands, is highly desirable. As elegantly demonstrated by Peter Dervan's group, sequence-selective recognition of successive G-stretches in the minor groove of B-DNA is enabled by netropsin (Net) and distamycin (Dist) derivatives in which at well-defined positions, a pyrrole ring is replaced by an imidazole. This enables the imidazole electron-rich N atom to H-bond in-register to the extracyclic $N_2$ atom of a targeted guanine (2) (3) (4) (5). Some derivatives forming dimers connected by a short 'hairpin' alkylamide connector, each monomer binding to one DNA strand, were endowed with subnanomolar DNA-binding affinities (3).

Our Laboratories have a long-standing interest in sequence-selective recognition of guanines by non-covalent targeting of $O_6/N_7$ in the major groove. The majority of sequence-selective DNA recognition and binding by proteins take place in this groove (6) (7). We previously designed two classes of intercalator-oligopeptide derivatives using both theoretical and experimental studies.
- In the first class, the anchoring intercalator is the anthraquinone ring of the anti-cancer drugs ametantrone (AMT) or mitoxantrone (MTX) (Figure 1a). Both 'carrier' arms are identically substituted by an oligopeptide, having an Arg or a Lys side-chain at a proper position to target $O_6/N_7$ of two successive G bases on the nearest strand (8) (9) (10) (11). The present work focuses on novel derivatives of this class.
- In the second class, the intercalator is a porphyrin with three cationic N-methylpyridinium ring and one dimethoxyphenyl ring symmetrically substituted with an arginine (12) (13) (14). Energy-minimization studies showed each Arg side-chain to bind to $O_6/N_7$ of two successive G bases upstream of the intercalation site of the nearest strand.
For both classes, the preferential intercalation site of the parent compounds AMT/ MTX (15) (16) (17) or N-methylpyridinium porphyrin (18) was demonstrated to be $d(CpG)_2$. In the major groove, a preferential binding of Lys/Arg side-chains to $O_6/N_7$ of guanine bases rather than to any of the other sites of the other three bases should take place if such side-chains are at an optimized vicinity to two successive bases on the closest strand upstream of the intercalation site. Thus both classes of compounds should have the palindromic hexanucleotide d(GG**C** **G**CC)$_2$ sequence as a privileged recognition site (a blank is added to mark an intercalation site). The theoretical predictions were supported by experimental measurements of the melting temperatures ($\Delta T_m$) of the drugs complexes with their targeted sequence versus competing DNA sequences, such as, in the porphyrin series, d(TA**C** **G**CG)$_2$ (13) (14) or, in the anthraquinone series (10), a 'random' sequence encompassing a CG sequence. Notably, in this series, the parent AMT drug stabilized by nearly equal $\Delta T_m$ amounts the targeted and the random sequences, in marked contrast to the Lys derivative which enabled a 12° larger stabilization of the targeted sequence than the random one (10).

We mention here two additional experimental results, which we deem relevant to the present work:

-In the porphyrin series, along with the mono-intercalating derivative, bis- and tris-intercalating derivatives were designed and synthesized. The latter compound was shown to target the ten base-pair palindromic sequence d(**C** **G**GG**C** **G**CCC **G**)$_2$. Experimentally, when complexed to this sequence, it enabled a greater increase of $\Delta T_m$ than the monointercalator and interfered with topoisomerase I unwinding at a three-fold lower concentration (14). Such results led us to accordingly design trisintercalating derivatives in the MTX series as well.

- Very recently, an AMT Lys-derivative recently synthesized in two of our Laboratories (11), proved to be active in cellular assays against tumor cells. While it was ten times less active than MTX against tumor cells, it was fifty times less toxic than it on healthy cells. This motivates the present study, in which we introduce novel MTX derivatives designed to further increase their DNA affinities, their GC-rich preferences and the length of the bound oligonucleotide. There are 2080 different hexamer sequences, and 524800 decamer sequences (4). Recognizing a unique hexamer or decamer sequence that is involved in pathogenesis paves the way to augmented inhibition of tumor cell growth without adversely increasing the inhibition of healthy cells.

We will focus on the complexes of d(GG**C** **G**CC)$_2$ with two further evolved anthraquinone mono-intercalators and of d(**C** **G**GG**C** **G**CCC **G**)$_2$ with one novel trisintercalator and resort to long-duration MD to monitor the interactions that enable sequence-selectivity. Since these systems are highly charged and dynamic, and surrounded by ions and polarizable water molecules, we resort to a reliable computational model accounting for the underlying effects of polarization and electron anisotropy, AMOEBA, recently parameterized for DNA and RNA on the basis of Quantum Chemistry and tested in several oligonucleotide simulations (19). We use theTinker-HP software (20) on GPU (21). This should enable to evaluate whether the in-register ligand-DNA interactions with $O_6/N_7$ in the major groove are long-lived.

At this stage it is appropriate to mention some contexts of the frequent occurrence of two GC-rich sequences. Along with the palindromic sequence d(GGCGCC)$_2$ on which the present study is focused, there is a non-palindromic sequence, d(GGCGGG).d(CCCGCC), with five G bases and one C base on one strand and one G base and five C bases on the other (denoted as 5G/1C-5C/1G). The palindromic sequence is encountered in oncogenes such as the human retinoblastoma gene (22) but also in retroviruses such as HIV-1 (23). An outstanding example of its involvement in retrovirus integration is provided by the complex of polymerase of HIV-1 with a double-stranded DNA template-primer having the sequence d(TGGCGCCCG).d(CGGGCGCC) starting from the N-terminal part of the primer (24) (25).

Worth mentioning is a related example, that of yeast Ty1 RNA, where the start sequence is AGCGCC, double-stranded with the GGCGCU stem, thus with one GC base-pair replaced by an AU one (26). It could be instructive if future studies unraveled related motifs in other, emerging retroviruses.

Although this list is far from limitative, other involvements of the d(GGCGCC)$_2$ sequence are their occurrence as recognition site for seven Type II restriction enzymes (27) and as a binding site for nuclear transcription factor complexes (28), as a Smad-binding element (29).

The non-palindromic sequence d(GGCGGG).d(CCCGCC) is encountered in several oncogenes, such as *c-myc* (30), (31)*, c-kit* (32) *, sp1* (33), as well as in the enhancer region of Sarcoma Virus 40 (SV40) (1).

Owing to the potential importance of targeting such a sequence as well, we address the following point in the last part of the present work. Since a stable in-register binding to $O_6/N_7$ of two successive G bases by a dialkylammonium entity is demonstrated by long-duration MD, we show that a modular replacement by another entity results into the recognition of another sequence of two base-pairs, such as two successive C bases on one strand. This led us to replace in one arm this entity by bisimidazole. The deprotonated N of each imidazole is oriented in such a way as to accept a proton from the extracyclic $N_4$ atom of a C base, in the same way as, in imidazole-pyrrole polyamides, the extracyclic $N_2$ atom of a G base in the minor groove (3).

Recently, a GC content above genomic average has also been identified in extrachromosomal (ec-DNA) and extrachromosomal circular DNA (eccDNA). These small DNA arrangements, which mostly generate from genetic instability events, are non-chromosomically inherited and their transmission causes intratumoral genetic heterogeneity and may lead to therapeutic resistance (34), (35). They embody Alu repeat sequences in which both the palindromic and the non-palindromic sequences are found (36).

Starting with the Lys derivative of AMT reported in (11), fifteen mono- and trisintercalating novel derivatives have been designed to target the d(GG**C** G**C**C)$_2$ sequence. This study first focuses on two mono-intercalating derivatives, denoted as **I** and **II**, which were found to enable for the most permanent and closest in-register binding to O$_6$/N$_7$ of G$_1$-G$_2$/G$_1'$-G$_2'$. Their chemical structures are given in Figure 1(b-e). Both have a dialkylammonium -(CH$_2$)$_2$-(NH$_2$)$^+$-(CH$_2$)$_2$-(NH$_2$)$^+$-CH$_2$CH$_3$ side-chain instead of the Lys side-chain. The two peptide main-chains of **I** are, similar to the Lys starting compound, connected to the AMT core by 1,2,3-triazole. **II** has two 'true' peptide backbones, with two Cys residues flanking their central dialkylammonium side-chains. As put forth in our earliest paper (8), this enables the formation of two disulfide bridges between the two main-chains, further stabilizing their anti-parallel β-sheet conformation, which in turn stabilizes the in-register binding to G$_1$-G$_2$/G$_1'$-G$_2'$. We further report MD analyses of the most promising trisintercalator derivative, **III** (Figure 1d) in its complex with sequence d(**C** GG**C** G**C**CC **G**)$_2$. It derives from **II** by substituting the end C atom of each dialkylammonium with a peptide backbone substituted with imidazole: this is inspired by the work of Dervan *et al.* in the imidazole/pyrrole polyamide minor groove binders (3). And as in our previous work in the porphyrin class of derivatives (14), the imidazole groups are connected by another amide to a 9-aminoacridine ring destined to intercalate at the two d(CpG)$_2$ sequences on both 5' and 3' ends. We finally consider compound **IV** (Figure 1e) in complex with the non-palindromic sequence d(GG**C** GGG).d(CCC **G**CC) which it is destined to target. It differs from **III** by the replacement of one dialkylammonium side-chain on the arm facing the 5C/1G strand, by a bis-imidazole: the deprotonated N of the first and second imidazoles are destined to target the extracyclic N$_4$ of C$_1'$ and C$_2'$, respectively.

## Computational procedure.

MD simulations use the AMOEBA polarizable multipole potential (37) coded in the Tinker-HP software (20) in its recent GPU implementation (21). We use the 2018 DNA parameter (38). The distributed multipoles of compounds **I-IV** were derived by a Stone analysis (39) on their molecular wave functions computed with the B3LYP DFT functional (40) (41) and a cc-pVDZ basis set (42) (43) (44) with the G09 (45) software.

The mono- and trisintercalated DNA conformations were constructed using the JUMNA software (46). **I** and **II** were complexed to the 18-mer oligonucleotide d(TACGTAGG**C GC**CTACGTA)$_2$. **III** was complexed to the 18-mer oligonucleotide d(CGTA**C GGGC GCCC G**TACG)$_2$. **IV** was complexed to the 18-mer d(CGTA**C GGGC GGGC G**TACG).d(CGTA**C GCCC GCCC G**TACG).

*Complexes of I and II.* The anthraquinone ring was first manually docked in the central intercalation site d(CpG)$_2$. This was followed by a first step of constrained energy minimization in the absence of water and counterions, DNA being held rigid. On each arm, the distance constraints (2.2 Å) were between one –(NH$_2$)$^+$- proton of the first alkylammonium group and O$_6$ of G$_2$/G$_2$', and between its second proton and N$_7$ of G$_2$/G$_2$'. Similar constraints were set between one –(NH$_2$)$^+$- proton of the second alkylammonium group and O$_6$ of G$_1$/G$_1$', and between its second proton and N$_7$ of G$_1$/G$_1$'. The β-sheet conformation was enforced by 2.2 Å distance constraints between the main chain C=O and NH groups of one dialkylammonium and the main-chain NH and the C=O of the other. To enable for, and to stabilize, a symmetrical binding to the two strands, we set two additional distance constraints of 3.2 Å. On each arm, these were between the N atom of the cationic -(NH$_2$)$^+$- group preceding the peptide main-chain and O$_1$ of the phosphate of the central intercalation site which faces it. This interaction was identified in our first energy-minimization study of MTX with oligonucleotides (17). Additional distance restraints of 3.2 Å were between the carbonyl O and the amide H atoms of the formamide moieties of the peptide backbones, to limit deviations from planarity during MD. The DNA-ligand complex was immersed in a bath of water molecules in a box of dimensions 51, 54 and 90 Å on the x, y, and z dimensions. Neutrality was ensured by placing 28 Na$^+$ cations in initial random positions. Periodic boundary conditions (PBC) were applied along with Smooth Particle Mesh Ewald (PME) (47). We used cutoff values of 12 Å and 9 Å for van der Waals and Ewald interactions, respectively.

Constrained energy-minimization was resumed, now also relaxing DNA, water and counterions. Molecular Dynamics was subsequently run. The bond lengths connecting the H atoms were constrained with the SHAKE algorithm (48). Equilibration was started by 50 K stepwise raises in temperature for a duration of 1 ns at constant volume, from 0 to 300 K. Production was then started at 300 K with a Bussi thermostat (49) and at constant pressure 1 atm. Coordinates were saved every 100 ps.

On the basis of preliminary runs, we were led to introduce four additional 3.2 Å restraints between $O_4(T)$ and $N_1(A)$ and between $N_3(T)$ and $N_6(A)$ of the two-end base-pairs, $T_1$-$A_{18'}$ and $T_{1'}$-$A_{18}$. These were necessary to prevent the 'fraying' of these base-pairs in the course of MD, and their inclusion has precedents in DNA MD simulations including those with polarizable potentials (19). Large amplitude motion of DNA and the solvent were observed during the initial phases of the production runs. These can cause an excess kinetic energy transfer to the bound ligand, which could destabilize and disrupt its complex with DNA. This led us to retain the intermolecular DNA-ligand and the ligand β-sheet restraints in the initial production phases as well, until 32 ns were reached when visual inspection of the trajectory showed that the DNA fluctuations were significantly damped. Unrestrained MD was then started, retaining, however, the restraints enforcing formamide planarity and the four restraints enforcing H-bonding between the two-end base-pairs. The choice of this two-staged production should be justified by the much longer duration of the 'unrestrained' phase with damped DNA fluctuations than of the restrained one with incompletely damped DNA: namely, over 100 ns as compared to 32 ns. Any long-lived DNA-ligand interaction in the unrestrained phase should thus be meaningful. This is a significantly different situation than in ligand-protein simulations, for which the ligand binding site is to a large extent 'shielded' from the solvent and undergoes much more limited fluctuations.

*Complexes of **III** and **IV**.* We have used a closely similar protocol for energy-minimization and MD as for **I** and **II**. For the starting energy minimizations with rigid DNA, each 9-aminoacridine was docked in its $d(CpG)_2$ intercalation site by distance restraints (4.0 Å) between $N_9$ and the $N_3$ atoms of the four bases. These restraints were removed in the next stage of energy-minimization. For the MD runs, in order to prevent base-fraying, three intermolecular distance restraints of 3.1 Å were set between $N_4(C)$ and $O_6(G)$, $O_3(C)$ and $N_1(G)$, and $O_2(C)$ and $N_2(G)$ of base-pairs $C_1$-$G_{18}$ and $C_{1'}$-$G_{18'}$. In the complex of **IV**, the initial ligand-DNA distance restraints on the C-rich strand were between the deprotonated N of the first imidazole of the bis-imidazole side-chain and $N_4$ of $C_{8'}$ and between that of the second imidazole and $N_4$ of $C_{7'}$.

**Results and discussion.**

*A. Complex of **I**.*

Figures 2a-c give a representation of the complex of **I** with its target oligonucleotide in the last frame of a 150 ns unconstrained MD trajectory. The interactions of each dialkylammonium arm with the two targeted guanines are shown separately. Figure 3a shows the free energy landscape (FEL) calculated from this trajectory. FEL is represented using two variables that reflect specific properties of the system and measure conformational variability: the radius of gyration and the root mean square deviation (RMSD) with respect to the average conformation. The Gibbs free energy is estimated from the probability distribution of sampled populations. The zero energy is at 0 kJ/mol, and corresponds to the lowest energy conformational state (dark blue). A local energy minimum is observed over a large free energy space (deep basin, dark blue) indicating that these conformational ensembles are very stable and undergo small variations during the simulation period (less than 0.4 Å for the RMSD and 0.2 Å for the radius of gyration). Figure 3b shows the probability distribution of the two H-bond distances between the CO and NH groups of the main-chain around the dialkylammonium side-chains which stabilize the anti-parallel β-sheet conformation, along with the two corresponding distances between the terminal CO group of one main-chain and the partly acidic CH proton of the triazole ring. Figure 3c reports the probability distribution of three distances: the first two ones are between the N atom of the -(NH$_2$)$^+$- group of the AMT chain and the anionic O of the phosphate of the central intercalation site which faces it. The third one is between this N atom in the arm facing the primed strand and N$_7$ of the guanine base of the intercalation site on this strand, 5' with respect to the phosphate. Figures 3d and 3e show the probability distribution of the distances between O$_6$ and N$_7$ and their closest -(NH$_2^+$)- proton in the following dimers: on the primed and unprimed strand of: G$_1$ and the second alkylammonium group (3d) and G$_2$ and the first alkylammonium (3e). The corresponding time series are also reported in Supp. Info Figure S1.

Figures 3d-e attest to the lasting stability of the bidentate complexes of each dialkylammonium with O$_6$/N$_7$ of its targeted guanine. Eight H-bond distances are considered. All four distances involving O$_6$ evolve in a range mostly in a 2.0-2.3 Å interval, with occasional increases to 2.4 Å. In three of the complexes, the distances involving N$_7$ are in the 2.15-2.4 Å interval. In one complex, that of the second dialkylammonium with G$_1$ on the unprimed strand, this distance is notably larger, in the range 2.6-3.0 Å, thus the 'loosest' of the eight distances stabilizing the four bidentate complexes. To which extent were the two 'central' backbone CO—HN H-

bonding distances preserved in the course of MD? Figure 3b show that the first of these distances was in a 2.2-2.6 Å interval, while the second underwent larger amplitudes with values reaching 3.0 Å and in ten structures, above this value. However, such loosenings/disruptions of the inter-backbone H-bond distances did not impair the anti-parallel arrangement of the two backbones nor the in-register binding of the side-chains to $G_1$-$G_2$/$G_{1'}$-$G_{2'}$. The distances between the triazole acidic proton on one main-chain and the terminal CO of the other main-chain are much larger (Figure 3b). They have average values of 3.0 and 4.6 Å for the first and the second one, respectively. Figure 3c show that the two ammonium N atoms of the AMT arm interact differently with $O_1$ of the central phosphate group it faces. In the primed strand, the ammonium group binds simultaneously to $O_1$ and to $N_7$ of the G base 5' to it, with limited amplitudes of distance variations. By contrast, on the unprimed strand, there are no direct interactions with $O_1$.

It is instructive at this stage to refer to the crystal structure of Smad5-MH1 complexed with the palindromic sequence d[$G_1T_2A_3T_4G_5G_6C_7G_8C_9C_{10}A_{11}T_{12}A_{13}C_{14}$]$_2$ (29). As is the case with the two arms of ligand **I**, both monomer chains C and D of Smad5 contribute equally to the binding, and the binding to the G bases takes place in the major groove and involves solely basic residues. Thus, Arg75 of chain C binds bidentate to $G_5$ on the unprimed strand and Lys82 can bind simultaneously to $G_6$ and $G_{7'}$ on the unprimed and primed strands, respectively. Conversely, Arg75 of chain D binds to $G_{10'}$ while its Lys82 binds simultaneously to $G_{9'}$ and $G_8$ on the primed and unprimed strands, respectively. Thus, on the other hand while in the protein complex Lys82 of each chain can in the major groove 'straddle' the two DNA strands, in the complex of **I** each arm is confined to one strand. It will be instructive to compare the evolutions of such distances with two 'true' peptide main-chains connected together by two disulfide bridges (see below).

A Fourier Transform Infrared spectroscopy study was reported on complexes of a tricationic biogenic amine, spermidine (Spd: ($NH_3^+$)-($CH_2$)$_3$-($NH_2^+$)-($CH_2$)$_4$-$NH_3^+$) with DNA (50). It resulted into a structural model in which the first ammonium and the central alkylammonium bridge $N_7$ of two successive purine bases on the same strand, namely G and A for the sequence that was considered. Demonstration of the binding of a dialkylammonium entity to two successive purines in the major groove lends support to the use of such an entity to target two successive G bases as in compounds **I-IV** of this study, recalling that the electrostatic potential exerted by $N_7$ of guanine on an incoming positive charge is much stronger than that of $N_7$ of adenine (51).

*B. Complex of **II**.*

Figure 4 represents the complex of **II** with the 18-mer oligonucleotide at the outcome of a 137 ns unrestrained MD, focusing on the unprimed and primed strands, respectively. The FEL reported in Figure 5a shows a high conformational stability of the complex along the MD simulation, less than 0.1 Å for both the RMSD and the radius of gyration compared to the average structure. The four CO—HN H-bonds maintaining the anti-parallel β-sheet conformation have been enforced by the two disulfide bridges. The probability distance distribution of the two H-bonds around the dialkylammonium side-chain, and the two H-bonds at the N- and C-termini are reported in Figure 5b. Even though all four G bases are targeted throughout the trajectory, the bidentate mode persists only with base $G_1$ on the unprimed strand. For the three other bases, it is either $O_6$ or $N_7$ that is bound by an ammonium hydrogen. For clarification we have chosen to plot the evolutions of the distances of both ammonium hydrogens, denoted as $H_a$ and $H_b$, with respect to $O_6$ and $N_7$. We denote by $N_a$ and $N_b$ the first and second N atoms of the dialkylammonium group. In the unprimed strand, these are reported in Figures 5c and 5e for the distances between $O_6$ and $N_7$ of a targeted base and both ammonium protons borne by $N_a$ or by $N_b$, namely in succession $O_6/N_7(G_1)$—$H_a/H_b(N_b)$ and $O_6/N_7(G_2)$—$H_a/H_b(N_a)$. In the primed strand, the evolutions of the corresponding distances are plotted in Figures 5d and 5f. The corresponding time series are also reported in Supplementary Information Figure S2.

In the unprimed strand, Figure 5c show that $O_6(G_1)$ and $N_7(G_1)$ are each H-bonded by one proton of $N_a$, namely $H_b$ and $H_a$ respectively. $N_7$ of $G_2$ is H-bonded by both $H_a$ and $H_b$ of $N_a$ (Figure 5e). In the primed strand, $N_7$ of $G_{1'}$ is solely H-bonded by $H_a$ of $N_{b'}$, $O_6$ being at average distance of 2.7 Å from this hydrogen. $H_b(N_{b'})$ is turned toward the solvent phase (Figure 5d). On the other hand, with base $G_{2'}$ it is now only $O_6$ that is H-bonded to one ammonium proton, $H_b(N_{a'})$ (Figure 5f).

Contrary to **I**, there were no direct anionic interactions between the cationic AMT nitrogen atoms and the central phosphates, but such interactions were mediated by water molecules (not shown).

Limiting the conformational freedom of the two main-chains of **II** could have partly prevented an optimal bidentate binding to $O_6/N_7$. In the next stage, we consider an extension of **II** by derivatizing each alkylammonium side-chain with a selected connector whose other end is linked to a 9-aminoacridine intercalator. This will result into an extension of the binding site from six to ten base-pairs. Could some properly chosen connectors enable to 'realign' the ammonium groups and possibly restore bidentate binding to $O_6/N_7$ of all four targeted guanines?

*C. Complex of **III**.*

Molecules able to target a unique six-base sequence would bind to one site out of 2080 (2). It is compelling to try and extend the number of base-pairs targeted to n=10 and, if successful, beyond. This could be attempted by extending the terminal methyl of each dialkylammonium with a structured peptide or peptide-like chain with appropriate side-chains poised to bind in-register to $O_6/N_7$ of guanine, $N_6/N_7$ of adenine, $O_4/C_5$ of thymine, or $N_4$ of cytosine (7). Work is in progress along these lines in our Laboratories.

We intend here to target the ten base-pair d(**C GGGC GCC G**)$_2$ palindromic sequence, as done in our preceding work (14), specifically, one sequence out of 524 800. It is noted that the last eight bases are involved in the complex of the HIV-1 LTR with the polymerase (25). We found it equally forceful to directly target the two d(CpG)$_2$ upstream and downstream by an intercalator such as 9-aminoacridine (9-AA) as could be with a peptide-related entity. This relies on the established preference of aminoacridine and related intercalators for this dinucleotide. The nature and length of the connector between the dialkylammonium and 9-AA should then be optimized. We have considered several connectors as reviewed in (52). We report here the results of a long-duration MD with one of the most efficient connectors found so far. Its selection is inspired by the results from Dervan's group on minor groove targeting by Net and Dist derivatives (2,3): we resort to an amide-imidazole motive which owing to its 'crescent-shaped' form, enables 9-AA to project into the intercalation site from the major groove site without significant distortion of the non-intercalated part of the DNA backbone or of the ligand arms.

Figure 6 shows the complex at the outcome of a 160 ns unconstrained MD. Figures 7 show a structural investigation of the stability of this complex.

The FEL reported in Figure 7a again shows a very high conformational stability of the complex along the MD simulation, less than 0.1 Å for both the RMSD and the radius of gyration compared to the average structure. There are two basins present. However, their < 0.025 Å RMSD separation is not meaningful. Figure 7b-f show the probability distributions of: (b) the four CO—HN bonds stabilizing the anti-parallel β-sheet backbone conformation and (e) the distances between the central ammonium N of the AMT carrier and $O_1$ of the central phosphate facing it; the H-bond and anionic interaction distances stabilizing it, namely (c, d) the H-bond distances between the two protons of each alkylammonium and $O_6/N_7$ of the targeted G base; (f) the distance between the HN of the imidazole connector of each arm and $N_7$ of the 5' G base of the 9-AA intercalation site on the same strand: the corresponding H-bond is similar to the one taking place in the minor groove in the Net or Dist complexes with $N_3$ of adenine or $O_2$ of thymine (4). The corresponding time series are also reported in Supplementary Information Figure S3.

All such interactions appear long-lasting over the course of unconstrained MD. Compound **III** is thus able to afford simultaneously and possibly in a mutually reinforcing manner: bidentate chelation of $O_6/N_7$ of each of the four G bases by one dialkylammonium; maintenance of the four H-bonds stabilizing the anti-parallel β-sheet conformation; binding of each central ammonium N of the AMT carrier to $O_1$ of the central phosphate facing it; and in addition, an H-bond between the protonated imidazole N to $N_7$ and $O_6$ of the intercalated G on the same strand. We have also performed MD simulations upon augmenting the ionic strength to 0.15M/l, namely with a total of 51 $Na^+$ and 23 $Cl^-$ atoms. This did not impact the intermolecular DNA-**III** distances.

The design of **III** is outlined in Figure 8a and its complementarity with the targeted 10-mer sequence is highlighted in Figure 8b.

*D. Complex of **IV**.*

Long-duration MD on the complexes of **I-III** have demonstrated the long-lived nature of in-register interactions of each dialkylammonium arm with $G_1$-$G_2$/$G_{1'}$-$G_{2'}$ on its nearest strand. Could the 'modular' nature of such interactions be leveraged by replacing, in a Lego-like fashion, one dialkylammonium side-chain by another entity destined to target two other bases than guanines? We have thus attempted to target another GC-rich sequence, in which the two upstream G bases on the primed strand, $G_{1'}$ and $G_{2'}$, are replaced by two C bases, and, conversely, the two downstream C bases on the unprimed strand, $C_5$ and $C_6$, are replaced by two G bases. Thus the targeted sequence is $d(G_1G_2\mathbf{C_3}\ \mathbf{G_4}G_5G_6).d(C_{1'}C_{2'}\mathbf{C_{3'}}\ \mathbf{G_{4'}}C_{5'}C_{6'})$. Such a sequence is encountered in a very recurrent fashion in oncogenes and retroviruses. A most striking example is provided in an early paper showing its repeated occurrence within short intervals, in the human c-Ha-*ras* gene (1). We are presently considering several electron-rich motives destined to accept, in the primed strand, a proton from the extracyclic $N_4$ group of $C_{1'}$ and $C_{2'}$, while retaining the dialkylammonium on the unprimed strand to target $G_1$-$G_2$. We present here results obtained with compound **IV**, which embodies two imidazole rings connected by one methylene. The deprotonated N atoms of the first and second imidazoles are destined to accept a proton from $N_4$ of $C_{2'}$ and $C_{1'}$, respectively: this is the analog of the H-bond interactions occurring in the minor groove between the extracyclic $N_2$ group of guanine and the deprotonated N atom of imidazole polyamides put forth in (3). Figures 9a-c represents the interactions taking place in the unprimed and the primed strands at the outcome of a 150ns unrestrained MD. The interactions on the unprimed strand (Figure 9b) are the same as those found with compound **I**, from which **IV** is derived. On the primed strand, the deprotonated N of each imidazole is H-bonded to $N_4$ of the C base it is destined to target (Figure 9c). The two central H-bond interactions stabilizing the β-sheet conformation of the two main-chains of **IV** are preserved. On the other hand, the protonated N atom of the carrier chain is now H-bonded to $N_7$ of the 5'G of the intercalation site on the primed strand, rather than to $O_1$ of the central phosphate group as occurs on the unprimed strand. These results indicate the modular nature of the ligand-DNA interactions that can take place in the major groove when the ligand main-chain has a well-defined structural motive.

**Conclusions and Perspectives.**

The palindromic sequence d(GG**CG**CC)$_2$ is a recurring motive in oncogenes (22) and retroviruses (23). In order to selectively target it, we have designed two novel oligopeptide derivatives of the antitumor intercalator drug mitoxantrone (MTX). In these, in line with our previous work (8-11), while the anthraquinone ring is intercalated in the central d(CpG)$_2$ step, MTX is extended by two identical oligopeptide arms. In both derivatives, the central side-chain is a dialkylammonium, $-CH_2CH_2(NH_2)^+CH_2CH_2(NH_2)^+CH_3$, destined to target in the major groove the two successive upstream G bases on the nearest strand. This differs from our previous work in which the targeting side-chain was that of Lys (8, 10, 11) or of Arg (9). The main-chain of the first compound, **I**, has a 1,2,3-triazole, as in our recent work (11), and that of the second compound, **II**, has two Cys residues on both sides of the cationic side-chain, as in a previous study (9). We have performed long-duration polarizable MD simulations benefitting from the recent adaptation of the Tinker-HP software on GPU computers. These bore on the complexes of **I** and **II** with an 18-mer double-stranded DNA having the palindromic sequence in its center. During the simulation, in both complexes the two peptide main-chains were stabilized in an anti-parallel β-sheet conformation. On both arms, the first and second $(NH_2)^+$ groups interacted with, respectively, $O_6$ and/or $N_7$ of the second and of the first guanine on the nearest DNA strand. This should ensure for selective in-register recognition of $G_1/G_{1'}$ and $G_2/G_{2'}$, since neither A, T, nor C bases could compete with G for the binding of a cationic entity in the major groove. In addition, the length of the selected dialkylammonium is too short for it to reach out to a G base on the farther strand. We have then extended both dialkylammonium side-chains of **II** by an amide-imidazole group connecting it to an additional intercalator, 9-aminoacridine (9-AA). The 'crescent-shaped' nature of the connector enabled each 9-AA to intercalate into an upstream $d(CpG)_2$ sequence. As in previous work on an Arg derivative of a tricationic porphyrin (14), compound **III** thus acts as a trisintercalator, destined to selectively recognize the 10-base pair sequence d(**C** GGGC GCCC **G**)$_2$ palindromic sequence, namely one site out of 524 800. Along with this previous contribution and the present one, there are earlier studies on the complexes of bis- (53-57) and tris- and poly-intercalators with DNA (58-61). A compelling example is provided with a bisintercalating derivative of daunorubicin, endowed with picomolar DNA binding affinity, while the mono-intercalating parent compound had submicromolar affinity (62). With the exception of ditercalinium (55, 56) and the threading intercalators reported in (61, 63, 64), the connector chains were in the minor groove. Acridine trimers studied by Laugaa *et al.* (59) were shown to have DNA affinities in the range of DNA regulatory proteins. This implies that compound **III** could accordingly be endowed with enhanced DNA affinity in addition to its sequence-selectivity.

The presence of the two disulfide bridges in the main-chain, and that of the crescent-shaped amide-imidazole connector between the dialkylammonium side-chain and 9-AA enabled on each strand the simultaneous onset of: a) bidentate binding of the first and second -$(NH_2)^+$- groups to $O_6/N_7$ of $G_2/G_{2'}$ and $G_1/G_{1'}$ respectively; b) ionic interactions between the ammonium group of the AMT carrier and the facing phosphate of the central intercalation site; c) an H-bond of the protonated imidazole with $N_7$ of the nearest 5'G base of the 9-AA intercalation site. Such interactions persisted throughout the entirety of long-duration MD, together with all four intramolecular CO-HN bonds stabilizing the antiparallel β-sheet conformation.

Recent procedures have been leveraged to search for alternative minima resorting to unsupervised adaptive sampling and clustering (65). We plan to apply these in forthcoming work. Nevertheless, the present structural results on the complexes of **I-III** are deemed stable and conclusive, with, notably**, III** fulfilling all the interactions it could contribute to DNA binding. It will be nevertheless instructive to search for alternative binding modes of **I-III** if any, and this will be reported subsequently.

We did not seek at this stage to investigate sequence-selective recognition, and focused instead on one single hexamer sequence targeted by **I** and **II**, and one single decamer sequence targeted by **III**. We considered that G bases have no competitors for major groove binding by dialkylammonium. The sole competing sequence would have two successive G's on the opposite strand, namely, d($C_1C_2$**$C_3$** **$G_4$**$G_5G_6$)$_2$: this could occur with a longer side-chain, such as Lys, which could reach out to $G_5$-$G_6$/$G_{5'}$-$G_{6'}$. It is noted in this connection that an AMT Lys derivative was experimentally found to interact with a larger affinity with a central d(**CCC GGG**)$_2$ than with the central d(**GGC GCC**)$_2$ sequence (11). The shortening of the distance between the main-chain $C_\alpha$ and the first ammonium group of **I-III** to two methylene groups (as compared to four methylenes in Lys) was done deliberately to enforce its proximity to $G_2$/$G_{2'}$ and prevent its 'wandering' to the opposite strand. We are presently comparing the structures of complexes of these two palindromic sequences with a series of AMT derivatives with dialkylammonium side-chains differing in their length, as well as with the more 'canonical' Lys, Orn, and Arg side-chains, and the results will be reported subsequently.

This is a true incentive for prospective absolute binding free energy (ABFE) calculations aiming to compare the affinities for a given DNA sequence of related ligands designed in the mono- and trisintercalating series. However, such calculations remain a daunting computational challenge (66) unlike ligand-protein or even protein-protein complexes which received a lot of attention during the last. Accurate estimation of standard DNA-ligand binding free energies are considerably more sensitive to the protocol adopted than in the case of small inhibitor-protein complexes especially when the ligand is a large charged intercalator with a highly flexible elongated structure. In this case, the ligand could have two orders of magnitude smaller molecular weights than the protein, while ligands such as **I-IV** have less than one order of magnitude smaller weights than the oligonucleotide. Furthermore, both the DNA target and the ligand remain exposed to the solvent and the counterions, in contrast to ligand-protein complexes in which the ligand is in a partly buried protein cavity and interacts with only a limited number of 'discrete' water molecules without intrusions from a counterion as occurred in some initial simulations on the oligonucleotide complexes. In addition, the presence of two arms on the ligand implies that a dual set of Boresch (67) ligand-DNA anchoring restraints, or possibly more, must be applied to maintain the ligand in place during these simulations. The sensitivity of the ABFE outcome to the actual magnitudes of these constraints may also have to be considered. Hence, all these different factors (configurational entropy change, loss of entropy arising from the introduction of geometric restraints, end-points refinements, sufficient sampling of the lambda windows ...) should be taken into account when developing a tailored protocol within a well-defined workflow for ABFE calculations of major groove binding intercalators. This work is ongoing and will make the subject of an upcoming paper.

We have extended this study to the recognition of a non-palindromic sequence, d($G_1G_2\mathbf{C_3}$ $\mathbf{G_4}G_5G_6$). d($C_{1'}C_{2'}\mathbf{C_{3'}}$ $\mathbf{G_{4'}}C_{5'}C_{6'}$), itself encountered in numerous oncogenes (30), (31), (32), (33) and retroviruses (1). We have thus designed compound **IV**, in which one dialkylammonium arm is replaced by two imidazoles connected by one methylene. The interactions of the electron-rich N atom of the first and second imidazole rings with the extracyclic $N_4$ of respectively $C_{2'}$ and $C_{3'}$ on the C-rich strand were found to persist during the entire long-duration MD. These occurred concomitantly with those of the dialkylammonium group with $G_1$ and $G_2$ on the G-rich strand, as in compound **I** whence **IV** was derived.

The most significant findings could be summarized as follows:

1. The lasting, in-register, binding of a dialkylammonium side-chain to $O_6/N_7$ of two successive guanines upstream of the central intercalation site, enabling compounds **I-III** to selectively target the palindromic sequence d(GGCGCC)$_2$;

2. The long-duration stabilization of the anti-parallel β-sheet conformation of the two main-chains. This is a rare motif for DNA recognition by major groove-binding proteins, which predominantly adopt the α-helical conformation. Its first experimental occurrence was for the DNA-Arc repressor complex (68), (69). Its stabilization was predicted in the first energy-minimizations performed on forerunners of the present compounds (8, 9). Much later MD simulations with the classical AMBER force-field did support its persistence, but were limited to the 2ns duration of the production run (10). The present study can now confirm, in the context of polarizable MD, its persistence over much longer simulation times.

3. The use of the side-chains to grow connectors to another intercalator upstream. Such connectors are amide-imidazole as in the present study, or a peptide as well as a β-peptide (work in progress). The last cases could be a rare instance of the use of a peptide side-chain to grow a peptide *main-chain*: and the latter could in turn enable to grow another side-chain, now directed toward the strand opposite to the 'starting' side-chain. This will be reported by us in a forthcoming paper.

4. The fulfilment of all H-bonding capabilities of compounds **I-IV**, which include for **III,** ancillary interactions of the imidazole of both connectors with a G base of the 9-AA intercalation site.

5. The demonstration that optimized in-register side-chain binding to the base it targets enables for modularity. Thus a simple replacement of one dialkylammonium side-chain on one strand by bis-imidazole enables recognition of two cytosine bases instead of two guanines, thus targeting the hexameric core of the c-kit oncogene, d(GGCGGG).d(CCCGCC) instead of the d(GGCGCC)$_2$ palindrome.

As perspectives, we would like to mention three further extensions of the present work.

-The first is the recognition of longer DNA sequences (n> 10). This could be attempted by two possible means. One is a replacement of the connector by ordered peptides or pseudo-peptides with appropriate side-chains poised for in-register major groove recognition of $O_6/N_7$ of G, $N_6/N_7$ of A, $N_4$ of C, and $O_4$/5-$CH_3$ of T (see, eg, (7)). The other is substitution of each added intercalator, such as 9-AA, by tailored peptide or pseudo-peptide substituents extending beyond the added intercalation site to target well-defined purine or pyrimidine bases.

-The second is the replacement of 9-AA by other intercalators. As mentioned previously (14), 9-AA could be protonated. While on the one hand, the increased affinity for d(CpG)$_2$ could be compensated by a corresponding adverse increase of the desolvation energy, on the other hand, it could result into an augmented preference for d(CpG)$_2$ over other Pyr-Pur sequences. This is due to the much more favorable electrostatic/polarization interactions of a protonated aromatic ring with G than with any other base. In addition, the presence of a cationic charge on a conjugated molecule could facilitate transport across membranes, as shown in the case of oligo-Arg lipopeptides (70). 9-AA could also be substituted by groups such as –$NHSO_2CH_3$ as in the case of the anti-cancer drug m-amsacrine.

The third concerns the possibility of sequence-selective DNA cleavage, which can be achieved by several means:

a) the first relates to the intrinsic redox properties of the anthraquinone ring;

b) the second is based on trisintercalation being a driving force for DNA binding. As a consequence, controlled shortenings of the connector in trisintercalating derivatives could produce kinks of the DNA along its axis, enhancing DNA sensitivity for cleavage by endonucleases;

-c) the third resorts to organo-metallic compounds, as illustrated in eg (71-76), either grafted to the additional intercalators, or replacing them altogether. There are precedents among minor-groove binders. Thus a sequence-specific DNA-cleaving molecule was designed which encompasses a copper-complexing peptide, a netropsin residue and an acridine (76).

The presence of GC-rich sequences in extrachromosomal DNA (ec-DNA), very recently identified as a factor responsible for therapeutic resistance (35, 36) is a clear additional motivation to attempt their targeting by theoretical computations in synergy with chemical synthesis and in vitro and in vivo tests.

**Acknowledgements.** We wish to thank the Grand Equipement de Calcul Intensif (GENCI): Institut du Développement et des Ressources en Informatique (IDRIS), Centre Informatique de l'Enseignement Supérieur (CINES), France, project x-2009-07509, and the Centre Régional Informatique et d'Applications Numériques de Normandie (CRIANN), project 19980853. This work has received funding from the European Research Council (ERC) under the European Union's Horizon 2020 research and innovation program (grant agreement No 810367), project EMC2 (JPP).

**References.**

1. Mattes, W.B., Hartley, J.A., Kohn, K.W. and Matheson, D.W. (1988) GC-rich regions in genomes as targets for DNA alkylation. *Carcinogenesis*, **9**, 2065-2072
2. Dervan, P.B. (1986) Design of sequence-specific DNA-binding molecules. *Science (New York, N.Y.)*, **232**, 464-471.
3. Bremer, R.E., Baird, E.E. and Dervan, P.B. (1998) Inhibition of major-groove-binding proteins by pyrrole-imidazole polyamides with an Arg-Pro-Arg positive patch. *Chemistry & Biology*, **5**, 119-133.
4. Dervan, P.B. (2001) Molecular recognition of DNA by small molecules. *Bioorganic & medicinal chemistry*, **9**, 2215-2235.
5. Fechter, E.J. and Dervan, P.B. (2003) Allosteric Inhibition of Protein−DNA Complexes by Polyamide−Intercalator Conjugates. *Journal of the American Chemical Society*, **125**, 8476-8485.
6. Steitz, T.A. (1990) Structural studies of protein-nucleic acid interaction: the sources of sequence-specific binding. *Q Rev Biophys*, **23**, 205-280.
7. Segal, D.J., Dreier, B., Beerli, R.R. and Barbas, C.F., 3rd. (1999) Toward controlling gene expression at will: selection and design of zinc finger domains recognizing each of the 5'-GNN-3' DNA target sequences. *Proc Natl Acad Sci U S A*, **96**, 2758-2763.
8. Gresh, N. and Kahn, P.H. (1990) Theoretical Design of Novel, 4 Base Pair Selective Derivatives of Mitoxantrone. *Journal of Biomolecular Structure and Dynamics*, **7**, 1141-1160.
9. Gresh, N. and Kahn, P.H. (1991) Theoretical design of a bistetrapeptide derivative of mitoxantrone targeted towards the double-stranded hexanucleotide sequence d(GGCGCC)2. *Journal of biomolecular structure & dynamics*, **8**, 827-846.
10. Gianoncelli, A., Basili, S., Scalabrin, M., Sosic, A., Moro, S., Zagotto, G., Palumbo, M., Gresh, N. and Gatto, B. (2010) Rational design, synthesis, and DNA binding properties of novel sequence-selective peptidyl congeners of ametantrone. *ChemMedChem*, **5**, 1080-1091.
11. Ongaro, A., Ribaudo, G., Braud, E., Ethève-Quelquejeu, M., De Franco, M., Garbay, C., Demange, L., Gresh, N. and Zagotto, G. (2020) Design and synthesis of a peptide derivative of ametantrone targeting the major groove of the d(GGCGCC)2 palindromic sequence. *New Journal of Chemistry*, **44**, 3624-3631.
12. Gresh, N. and Perrée-Fauvet, M. (1999) Major versus minor groove DNA binding of a bisarginylporphyrin hybrid molecule: A molecular mechanics investigation. *Journal of Computer-Aided Molecular Design*, **13**, 123-137.
13. Mohammadi, S., Perrée-Fauvet, M., Gresh, N., Hillairet, K. and Taillandier, E. (1998) Joint molecular modeling and spectroscopic studies of DNA complexes of a bis(arginyl) conjugate of a tricationic porphyrin designed to target the major groove. *Biochemistry*, **37**, 6165-6178.
14. Far, S., Kossanyi, A., Verchère-Béaur, C., Gresh, N., Taillandier, E. and Perrée-Fauvet, M. (2004) Bis- and Tris-DNA Intercalating Porphyrins Designed to Target the Major Groove: Synthesis of Acridylbis-arginyl-porphyrins, Molecular Modelling of Their DNA Complexes, and Experimental Tests. *European Journal of Organic Chemistry*, **2004**, 1781-1797.
15. Lown, J.W., Morgan, A.R., Yen, S.F., Wang, Y.H. and Wilson, W.D. (1985) Characteristics of the binding of the anticancer agents mitoxantrone and ametantrone and related structures to deoxyribonucleic acids. *Biochemistry*, **24**, 4028-4035.
16. Lown, J.W. and Hanstock, C.C. (1985) High Field 1H-NMR Analysis of the 1:1 Intercalation Complex of the Antitumor Agent Mitoxantrone and the DNA Duplex [d(CpGpCpG)]2. *Journal of Biomolecular Structure and Dynamics*, **2**, 1097-1106.
17. Chen, K.X., Gresh, N. and Pullman, B. (1986) A theoretical investigation on the sequence selective binding of mitoxantrone to double-stranded tetranucleotides. *Nucleic Acids Res*, **14**, 3799-3812.
18. Pasternack, R.F., Gibbs, E.J. and Villafranca, J.J. (1983) Interactions of porphyrins with nucleic acids. *Biochemistry*, **22**, 5409-5417.
19. Zhang, C., Lu, C., Jing, Z., Wu, C., Piquemal, J.-P., Ponder, J.W. and Ren, P. (2018) AMOEBA Polarizable Atomic Multipole Force Field for Nucleic Acids. *Journal of Chemical Theory and Computation*, **14**, 2084-2108.
20. Lagardère, L., Jolly, L.-H., Lipparini, F., Aviat, F., Stamm, B., Jing, Z.F., Harger, M., Torabifard, H., Cisneros, G.A., Schnieders, M.J. *et al.* (2018) Tinker-HP: a massively parallel molecular dynamics package for multiscale simulations of large complex systems with advanced point dipole polarizable force fields. *Chemical Science*, **9**, 956-972.
21. Adjoua, O., Lagardère, L., Jolly, L.-H., Durocher, A., Very, T., Dupays, I., Wang, Z., Inizan, T.J., Célerse, F., Ren, P. *et al.* (2021) Tinker-HP: Accelerating Molecular Dynamics Simulations of Large Complex Systems with Advanced Point Dipole Polarizable Force Fields Using GPUs and Multi-GPU Systems. *Journal of Chemical Theory and Computation*, **17**, 2034-2053.
22. Hong, F.D., Huang, H.J., To, H., Young, L.J., Oro, A., Bookstein, R., Lee, E.Y. and Lee, W.H. (1989) Structure of the human retinoblastoma gene. *Proceedings of the National Academy of Sciences of the United States of America*, **86**, 5502-5506.
23. Wain-Hobson, S., Sonigo, P., Danos, O., Cole, S. and Alizon, M. (1985) Nucleotide sequence of the AIDS virus, LAV. *Cell*, **40**, 9-17.
24. Jacobo-Molina, A., Ding, J., Nanni, R.G., Clark, A.D., Lu, X., Tantillo, C., Williams, R.L., Kamer, G., Ferris, A.L. and Clark, P. (1993) Crystal structure of human immunodeficiency virus type 1 reverse transcriptase complexed with double-stranded DNA at 3.0 A resolution shows bent DNA. *Proceedings of the National Academy of Sciences*, **90**, 6320-6324.
25. Ding, J., Das, K., Hsiou, Y., Sarafianos, S.G., Clark, A.D., Jacobo-Molina, A., Tantillo, C., Hughes, S.H. and Arnold, E. (1998) Structure and functional implications of the polymerase active site region in a complex of HIV-1 RT with a double-stranded DNA template-primer and an antibody fab fragment at 2.8 Å resolution11Edited by J. Karn. *Journal of Molecular Biology*, **284**, 1095-1111.
26. Friant, S., Heyman, T., Wilhelm, M.L. and Wilhelm, F.X. (1996) Extended interactions between the primer tRNAi(Met) and genomic RNA of the yeast Ty1 retrotransposon. *Nucleic acids research*, **24**, 441-449.
27. Gowers, D.M., Bellamy, S.R.W. and Halford, S.E. (2004) One recognition sequence, seven restriction enzymes, five reaction mechanisms. *Nucleic Acids Research*, **32**, 3469-3479.
28. Korchynski, O., and ten Dijke, P. (2002) Identification and Functional Characterization of Distinct Critically Important Bone Morphogenetic Protein-specific Response Elements in the Id1 Promoter. *Journal of Biological Chemistry*, **277**, 4883-4891.
29. Chai, N., Li, W.-X., Wang, J., Wang, Z.-X., Yang, S.-M. and Wu, J.-W. (2017) Structural basis for the Smad5 MH1 domain to recognize different DNA sequences. *Nucleic acids research*, **45**, 6255-6257.
30. Seeburg, P.H., Colby, W.W., Capon, D.J., Goeddel, D.V. and Levinson, A.D. (1984) Biological properties of human c-Ha-ras1 genes mutated at codon 12. *Nature*, **312**, 71-75.
31. Jun, S., Hiroo, M., Hatsumi, T., Xiaoren, T., Masatoshi, M., Keiichi, I., Ichirou, K., Kailai, S. and Kazunari, K.Y. (1998) Genomic Organization and Expression of a Human Gene for Myc-associated Zinc Finger Protein (MAZ)*. *Journal of Biological Chemistry*, **273**, 20603-20614.
32. Yarden, Y., Kuang, W.J., Yang-Feng, T., Coussens, L., Munemitsu, S., Dull, T.J., Chen, E., Schlessinger, J., Francke, U. and Ullrich, A. (1987) Human proto-oncogene c-kit: a new cell surface receptor tyrosine kinase for an unidentified ligand. *The EMBO journal*, **6**, 3341-3351.
33. Kadonaga, J.T., Carner, K.R., Masiarz, F.R. and Tjian, R. (1987) Isolation of cDNA encoding transcription factor Sp1 and functional analysis of the DNA binding domain. *Cell*, **51**, 1079-1090.
34. Ain, Q., Schmeer, C., Wengerodt, D., Witte, O.W. and Kretz, A. (2020) Extrachromosomal Circular DNA: Current Knowledge and Implications for CNS Aging and Neurodegeneration. *International Journal of Molecular Sciences*, **21**, 2477.
35. Wildschutte, J.H., Baron, A., Diroff, N.M. and Kidd, J.M. (2015) Discovery and characterization of Alu repeat sequences via precise local read assembly. *Nucleic Acids Research*, **43**, 10292-10307.
36. Wu, S., Bafna, V., and Mischel, P. S. (2021) Extrachromosomal DNA (ecDNA) in cancer pathogenesis. *Current Opinion in Genetics & Development*, **66,** 78-82.
37. Ren, P., Wu, C. and Ponder, J.W. (2011) Polarizable Atomic Multipole-Based Molecular Mechanics for Organic Molecules. *Journal of Chemical Theory and Computation*, **7**, 3143-3161.
38. Zhang, C., Lu, C., Jing, Z., Wu, C., Piquemal, J.P., Ponder, J.W. and Ren, P. (2018) AMOEBA Polarizable Atomic Multipole Force Field for Nucleic Acids. *J Chem Theory Comput*, **14**, 2084-2108.
39. Stone, A.J. (2005) Distributed Multipole Analysis: Stability for Large Basis Sets. *Journal of Chemical Theory and Computation*, **1**, 1128-1132.
40. Lee, C., Yang, W. and Parr, R.G. (1988) Development of the Colle-Salvetti correlation-energy formula into a functional of the electron density. *Physical Review B*, **37**, 785-789.
41. Becke, A.D. (1993) A new mixing of Hartree–Fock and local density-functional theories. *The Journal of Chemical Physics*, **98**, 1372-1377.

42. Dunning, T. (1989) Gaussian-Basis Sets for Use in Correlated Molecular Calculations. 1. The Atoms Boron Through Neon and Hydrogen. *J. Chem. Phys.*, **90**, 1007-1023.
43. Dunning, Jr., T.H.D. (1989) Gaussian basis sets for use in correlated molecular calculations. I. The atoms boron through neon and hydrogen. *The Journal of Chemical Physics*, **90**, 1007-1023.
44. Feller, D. (1996) The role of databases in support of computational chemistry calculations. *Journal of Computational Chemistry*, **17**, 1571-1586.
45. Gaussian 09, Revision A.02, M. J. Frisch, G. W. Trucks, H. B. Schlegel, G. E. Scuseria, M. A. Robb, J. R. Cheeseman, G. Scalmani, V. Barone, G. A. Petersson, H. Nakatsuji, X. Li, M. Caricato, A. Marenich, J. Bloino, B. G. Janesko, R. Gomperts, B. Mennucci, H. P. Hratchian, J. V. Ortiz, A. F. Izmaylov, J. L. Sonnenberg, D. Williams-Young, F. Ding, F. Lipparini, F. Egidi, J. Goings, B. Peng, A. Petrone, T. Henderson, D. Ranasinghe, V. G. Zakrzewski, J. Gao, N. Rega, G. Zheng, W. Liang, M. Hada, M. Ehara, K. Toyota, R. Fukuda, J. Hasegawa, M. Ishida, T. Nakajima, Y. Honda, O. Kitao, H. Nakai, T. Vreven, K. Throssell, J. A. Montgomery, Jr., J. E. Peralta, F. Ogliaro, M. Bearpark, J. J. Heyd, E. Brothers, K. N. Kudin, V. N. Staroverov, T. Keith, R. Kobayashi, J. Normand, K. Raghavachari, A. Rendell, J. C. Burant, S. S. Iyengar, J. Tomasi, M. Cossi, J. M. Millam, M. Klene, C. Adamo, R. Cammi, J. W. Ochterski, R. L. Martin, K. Morokuma, O. Farkas, J. B. Foresman, and D. J. Fox, Gaussian, Inc., Wallingford CT, 2016.
46. Lavery, R., Zakrzewska, K. and Sklenar, H. (1995) JUMNA (junction minimisation of nucleic acids). *Computer Physics Communications*, **91**, 135-158.
47. Darden, T., York, D. and Pedersen, L. (1993) Particle mesh Ewald: An N·log(N) method for Ewald sums in large systems. *The Journal of Chemical Physics*, **98**, 10089-10092.
48. Jean-Paul, R., Giovanni, C. and Herman, J.C.B. (1977) Numerical integration of the cartesian equations of motion of a system with constraints: molecular dynamics of n-alkanes. *Journal of Computational Physics*, **23**, 327-341.
49. Bussi, G., Donadio, D. and Parrinello, M. (2007) Canonical sampling through velocity rescaling. *The Journal of Chemical Physics*, **126**, 014101.
50. Ouameur, A.-A., and Tajmir-Riahi, A.-H. (2004) Structural Analysis of DNA Interactions with Biogenic Polyamines and Cobalt(III) hexamine Studied by Fourier Transform Infrared and Capillary Electrophoresis. *The Journal of Biological Chemistry,* **279**, 42041-42054.
51. Pullman, B., and Pullman, A. (1981) Molecular Electrostatic Potential of the Nucleic Acids. *Quarterly Review of Biophysics,* **14,** 289-380.
52. Cheng, R.P., Gellman, S.H. and DeGrado, W.F. (2001) beta-Peptides: from structure to function. *Chem Rev*, **101**, 3219-3232.
53. Wakelin, L.P.G., Romanos, M., Chen, T.K., Glaubiger, D., Canellakis, E.S. and Waring, M.J. (1978) Structural limitations on the bifunctional intercalation of diacridines into DNA. *Biochemistry*, **17**, 5057-506.
54. Helbecque, N., Bernier, J.L. and Hénichart, J.P. (1985) Design of a new DNA-polyintercalating drug, a bisacridinyl peptidic analogue of Triostin A. *Biochemical Journal*, **225**, 829-832.
55. Lambert, B., Roques, B.P. and Le Pecq, J.B. (1988) Induction of an abortive and futile DNA repair process in E. coli by the antitumor DNA bifunctional intercalator, ditercalinium: role in polA in death induction. *Nucleic Acids Res*, **16**, 1063-1078.
56. Carrasco, C., Rosu, F., Gabelica, V., Houssier, C., De Pauw, E., Garbay-Jaureguiberry, C., Roques, B., Wilson, W.D., Chaires, J.B., Waring, M.J. *et al.* (2002) Tight Binding of the Antitumor Drug Ditercalinium to Quadruplex DNA. *ChemBioChem*, **3**, 1235-1241.

57. Hu, G.G., Shui, X., Leng, F., Priebe, W., Chaires, J.B. and Williams, L.D. (1997) Structure of a DNA-bisdaunomycin complex. *Biochemistry*, **36**, 5940-5946.
58. Gaugain, B., Markovits, J., Le Pecq, J.B. and Roques, B.P. (1984) DNA polyintercalation: comparison of DNA binding properties of an acridine dimer and trimer. *FEBS letters*, **169**, 123-126.
59. Laugâa, P., Markovits, J., Delbarre, A., Le Pecq, J.B. and Roques, B.P. (1985) DNA tris-intercalation: first acridine trimer with DNA affinity in the range of DNA regulatory proteins. Kinetic studies. *Biochemistry*, **24**, 5567-5575.
60. Wirth, M., Buchardt, O., Koch, T., Nielsen, P.E. and Norden, B. (1988) Interactions between DNA and mono-, bis-, tris-, tetrakis-, and hexakis(aminoacridines). A linear and circular dichroism, electric orientation relaxation, viscometry, and equilibrium study. *Journal of the American Chemical Society*, **110**, 932-939.
61. Vladimir, G., Jeeyeon, L., Jonathan, W., Steven, S., David, W.H. and Brent, L.I. (2001) Peptide bis-intercalator binds DNA via threading mode with sequence specific contacts in the major groove. *Chemistry & Biology*, **8**, 415-425.
62. Portugal, J., Derek J., Cashman, D. J., Trent, J. O., Ferrer-Miralles, N., Przewloka, T., Fokt, I., Priebe, W., and Chaires, J. B. (2005) A New Bisintercalating Anthracycline with Picomolar DNA Binding Affinity. *J. Med. Chem.*, **48**, 8209-8219.
63. Meredith, M.M., Matthew, T.H., Vladimir, G., Jinsong, R., Jonathan, B.C. and Brent, L.I. (2001) An octakis-intercalating molecule. *Bioorganic & medicinal chemistry*, **9**, 1141-1148.
64. Rhoden Smith, A. and Iverson, B.L. (2013) Threading Polyintercalators with Extremely Slow Dissociation Rates and Extended DNA Binding Sites. *Journal of the American Chemical Society*, **135**, 12783-12789.
65. Jaffrelot Inizan, T., Célerse, F., Adjoua, O., El Ahdab, D., Jolly, L.-H., Liu, C., Ren, P., Montes, M., Lagarde, N., Lagardère, L. *et al.* (2021) High-resolution mining of the SARS-CoV-2 main protease conformational space: supercomputer-driven unsupervised adaptive sampling. *Chemical Science*, **12**, 4889-4907.
66. Zhang, H., Gattuso, H., Dumont E., Cai, W., Monari, A., Chipot, C., Dehez, F. (2018) Accurate Estimation of the Standard Binding Free Energy of Netropsin with DNA. *Molecules* **23,** 228, doi:10.3390.
67. Boresch, S., Tettinger, F., and Leitgeb, M., Karplus, M. (2003) Absolute Binding Free Energies: A Quantitative Approach for Their Calculation. *J. Phys. Chem. B*, **107**, 9535–9551.
68. Zagorski, M.; Bowie, J.; Vershon, A.; Sauer, R.; Patel, D. J. (1989) NMR studies of Arc repressor mutants: proton assignments, secondary structure, and long-range contacts for the thermostable proline-8.fwdarw. leucine variant of Arc. *Biochemistry,* **28**, 9813-9825.
69. Breg, J.; Boelens, R.; George, V.; Kaptein, R. (1989) Sequence-specific proton NMR assignment and secondary structure of the Arc repressor of bacteriophage P22, as determined by two-dimensional proton NMR spectroscopy. *Biochemistry* **28**, 9826-9833.
70. Swiecicki, J.M., Di Pisa, M., Lippi, F., Chwetzoff, S., Mansuy, C., Trugnan, G., Chassaing, G., Lavielle, S. and Burlina, F. (2015) Unsaturated acyl chains dramatically enhanced cellular uptake by direct translocation of a minimalist oligo-arginine lipopeptide. *Chemical Communications*, **51**, 14656-14659.
71. Satyanarayana, S., Dabrowiak, J.C. and Chaires, J.B. (1993) Tris(phenanthroline)ruthenium(II) enantiomer interactions with DNA: Mode and specificity of binding. *Biochemistry*, **32**, 2573-2584
72. Geneviève, P., Jean, B. and Bernard, M. (1998) DNA And RNA Cleavage by Metal Complexes. *Advances in Inorganic Chemistry*, **45**, 251-312.

73. Gasser, G. and Metzler-Nolte, N. (2012) The potential of organometallic complexes in medicinal chemistry. *Current opinion in chemical biology*, **16**, 84-91
74. Lameijer, L.N., Ernst, D., Hopkins, S.L., Meijer, M.S., Askes, S.H.C., Le Dévédec, S.E. and Bonnet, S. (2017) A Red-Light-Activated Ruthenium-Caged NAMPT Inhibitor Remains Phototoxic in Hypoxic Cancer Cells. *Angewandte Chemie International Edition*, **56**, 11549-11553.
75. Boynton, A.N., Marcélis, L., McConnell, A.J. and Barton, J.K. (2017) A Ruthenium(II) Complex as a Luminescent Probe for DNA Mismatches and Abasic Sites. *Inorganic Chemistry*, **56**, 8381-8389.
76. Bailly, C., Sun, J.S., Colson, P., Houssier, C., Helene, C., Waring, M.J. and Henichart, J.P. (1992) Design of a sequence-specific DNA-cleaving molecule which conjugates a copper-chelating peptide, a netropsin residue and an acridine chromophore. *Bioconjugate Chemistry*, **3**, 100-103.

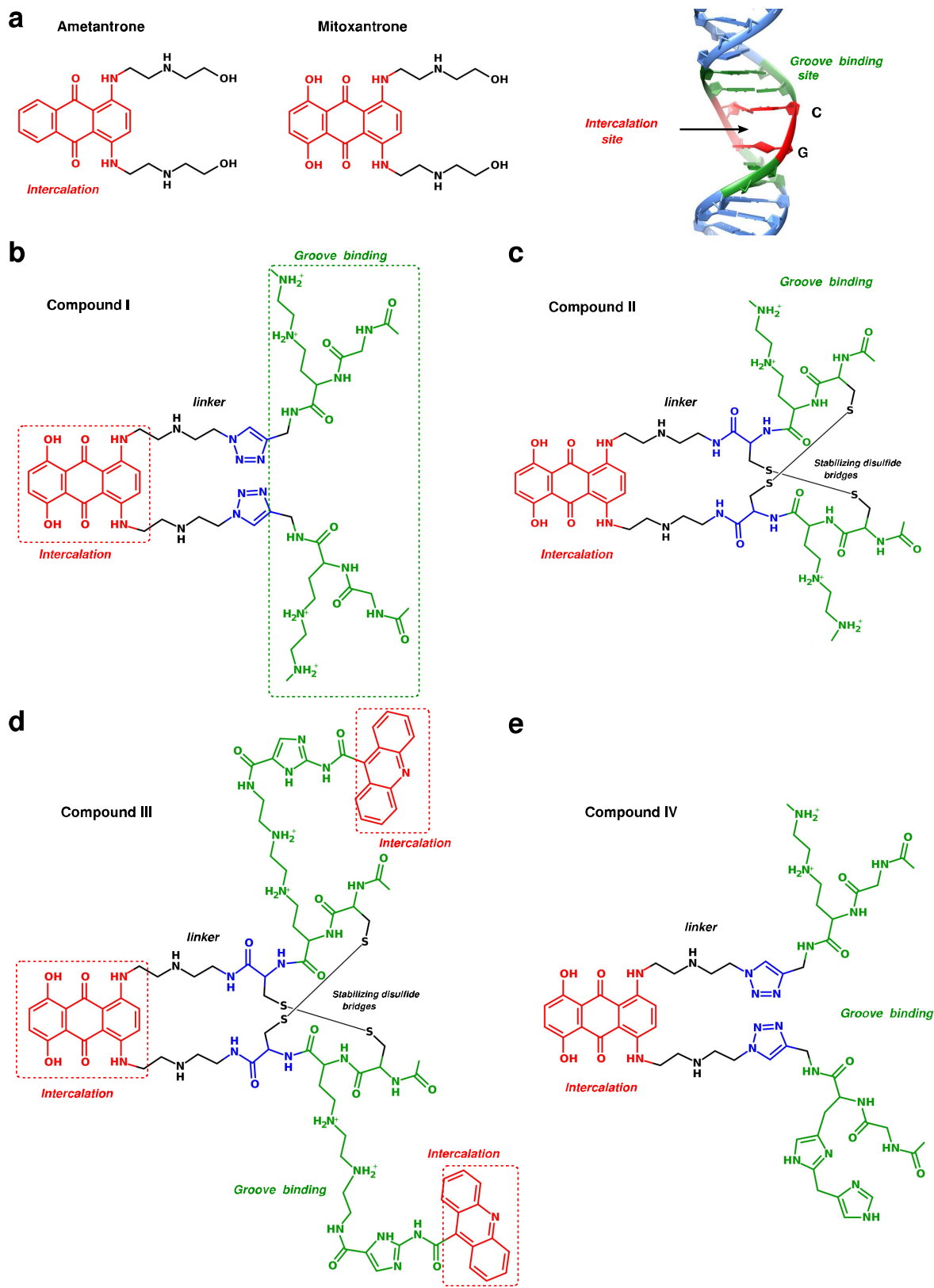

**Figure 1.** Molecular formulas of DNA intercalators. **a)** Anticancer drugs AMT (left) and MTX (middle). Right panel: Cartoon representation of a double stranded DNA with the CG intercalation site (*red*) and of the targeted groove binding region (*green*). **b-e)** Novel MTX derivatives: monointercalators (compounds **I**, **II**, and **IV**) and trisintercalator (compound **III**). Intercalation site, *red and red dashed box*; Groove binding site, *green and green dashed box*.

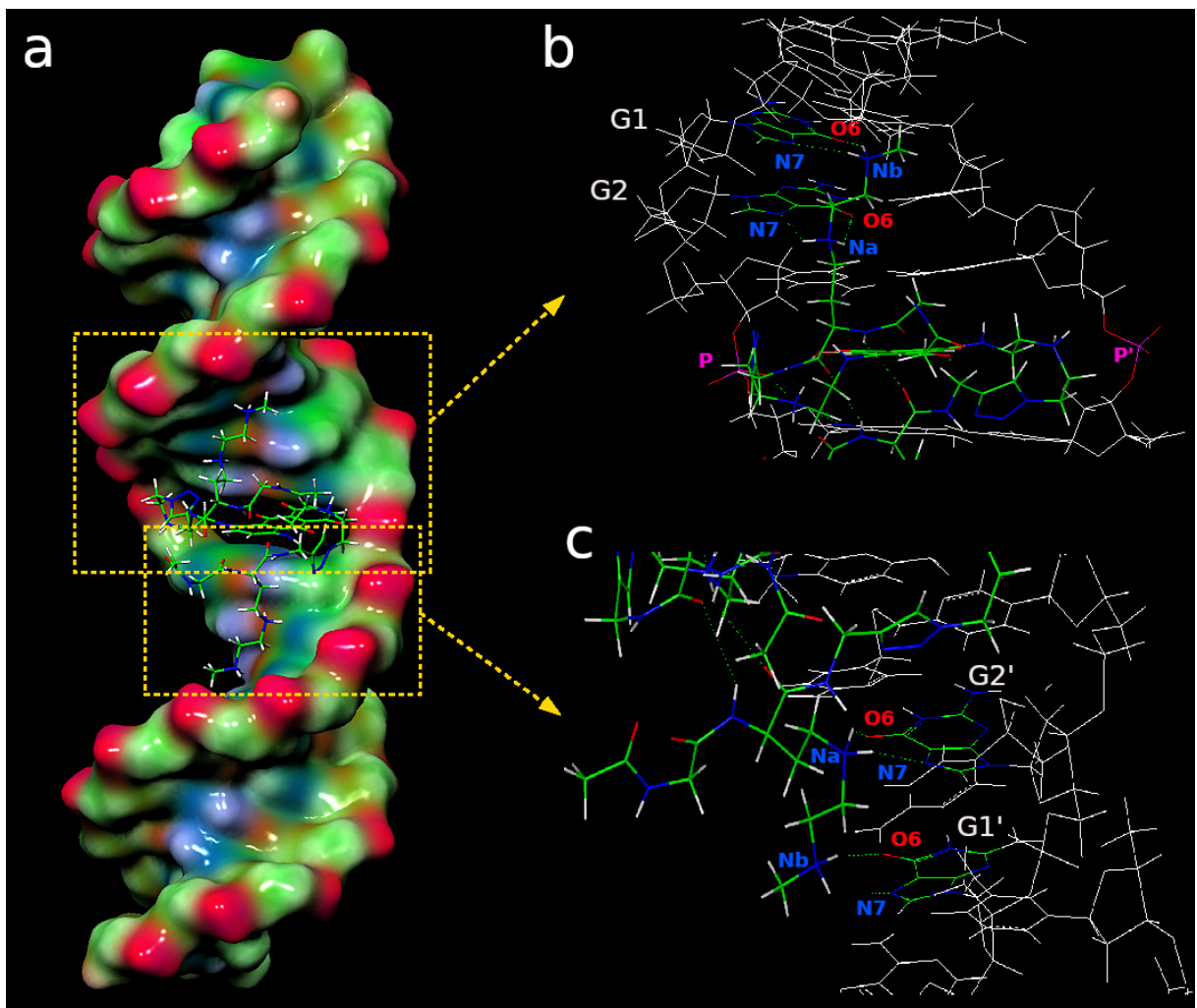

**Figure 2.** Compound **I** in complex with d(GG**C** **G**CC)₂ in the 18-mer oligonucleotide. **a)** overall view of the complex. Zoom on the upper **(b)** and lower **(c)** groove binding chains. Relevant atoms and nucleic bases are indicated.

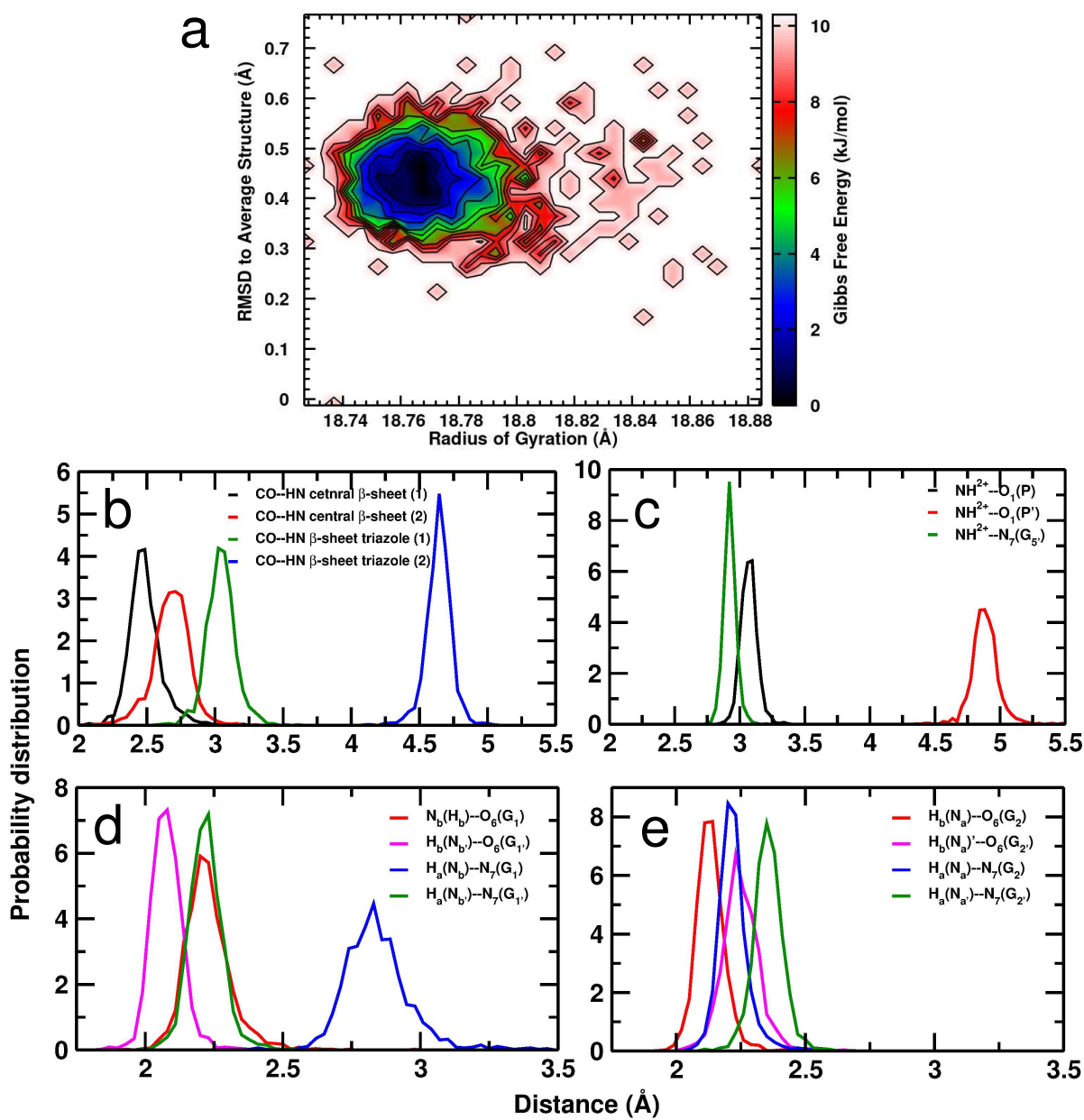

**Figure 3.** Structural investigation of **I** complexed with the target 18-mer oligonucleotide. **a)** Free energy landscape of the complex along the MD simulations; the zero energy is at 0 kcal/mol and corresponds to the lowest energy conformational state. **b-e)** Probability distribution of the following characteristic distances between **I** and the target: **(b)** β-sheet stabilizing main-chain CO—HN distances around the dialkylammonium side-chain and CO-HC(triazole); **(c)** ammonium N of the AMT carrier and $O_1$ of the central phosphate group facing it. The corresponding distance evolutions between the ammonium N on the primed site and $N_7$ of the 5' guanine of the intercalation site is also reported; **(d)** $O_6/N_7(G_1/G_{1'})$ with $H_a/H_b(N_b)$; $O_6/N_7(G_2/G_{2'})$ with $H_a/H_b(N_a/N_{a'})$ **(e)**. The corresponding time series are also reported in Supplementary Information Figure S1.

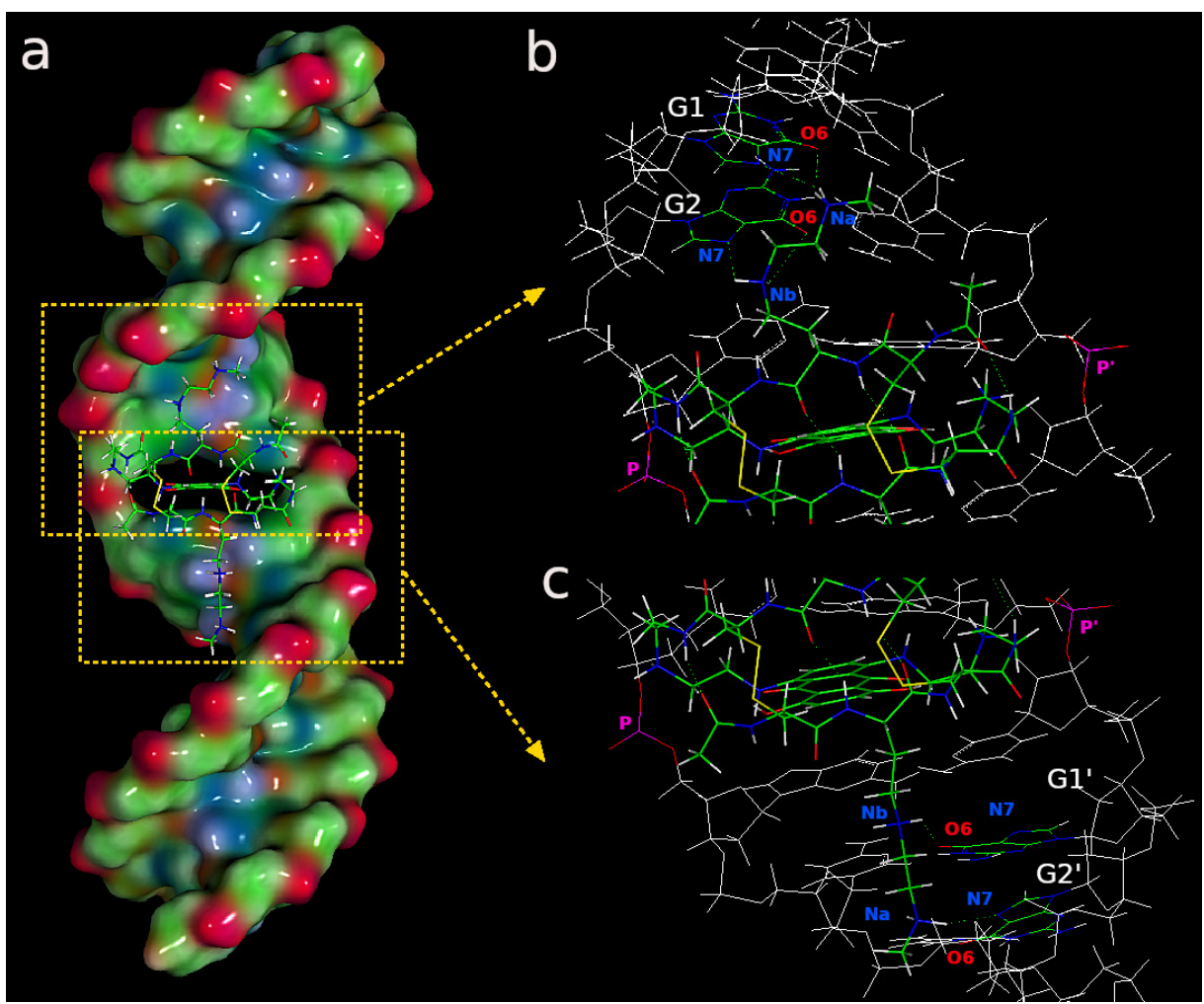

**Figure 4.** Compound **II** in complex with d(**GGC G**CC)₂ in the 18-mer oligonucleotide. **a)** overall view of the complex. Zoom on the upper **(b)** and lower **(c)** groove binding chains. Relevant atoms and nucleic bases are indicated.

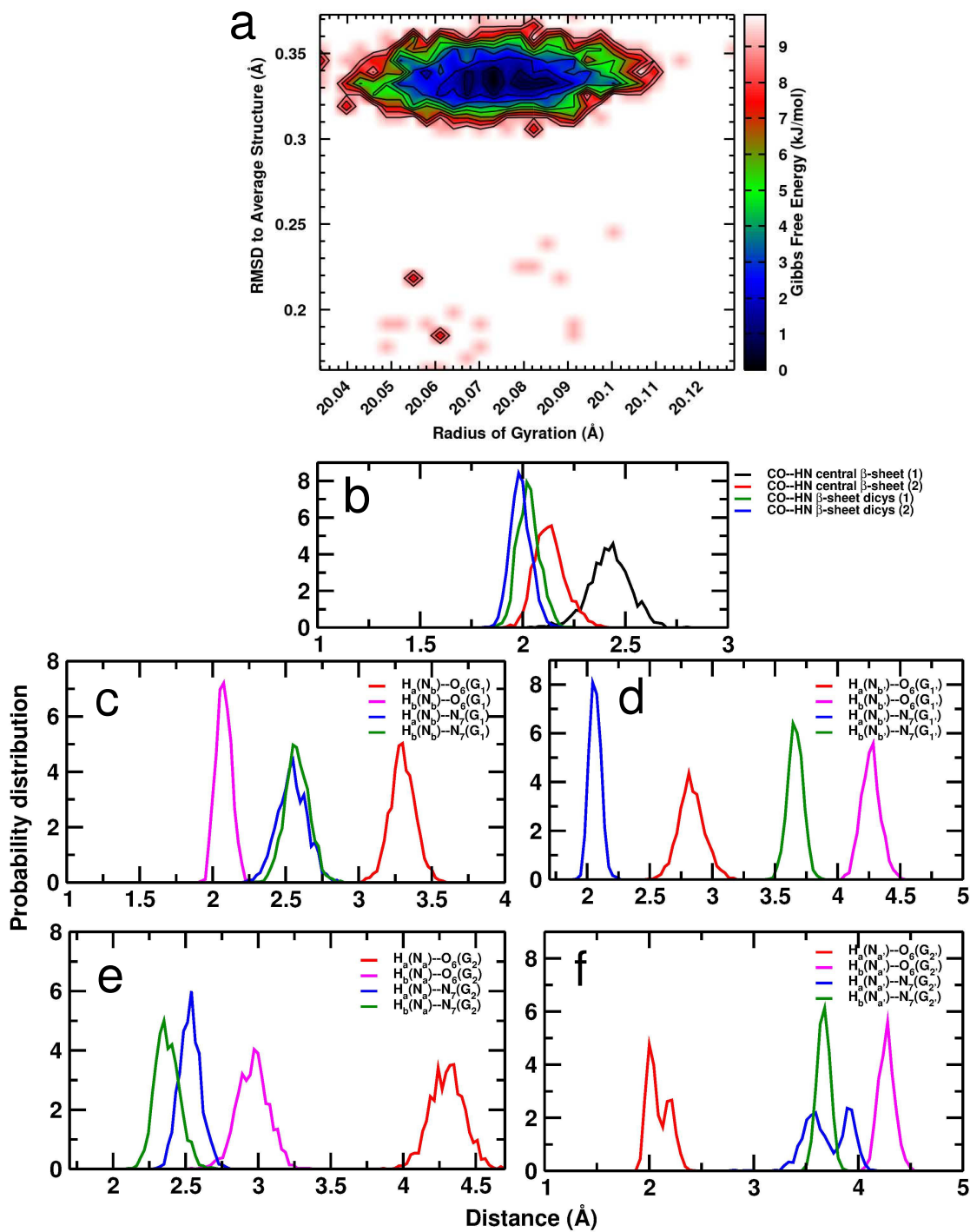

**Figure 5.** Structural investigation of **II** complexed with the target 18-mer oligonucleotide. **a)** Free energy landscape of the complex along the MD simulations; the zero energy is at 0 kcal/mol and corresponds to the lowest energy conformational state. **b-f)** Probability distribution of the following characteristic distances between **II** and the target: **(b)** β−sheet stabilizing main-chain CO—HN distances around the dialkylammonium side-chain and in the N- and C-termini; **(c)** $O_6/N_7(G_1)$ with $H_a/H_b$ ($N_b$); **(d)** $O_6/N_7(G_{1'})$ with $H_a/H_b$ ($N_{b'}$); **(e)**

$O_6/N_7(G_2)$ with $H_a/H_b(N_a)$; **(f)** $O_6/N_7(G_{2'})$ with $H_a/H_b(N_{a'})$. The corresponding time series are also reported in Supplementary Information Figure S2.

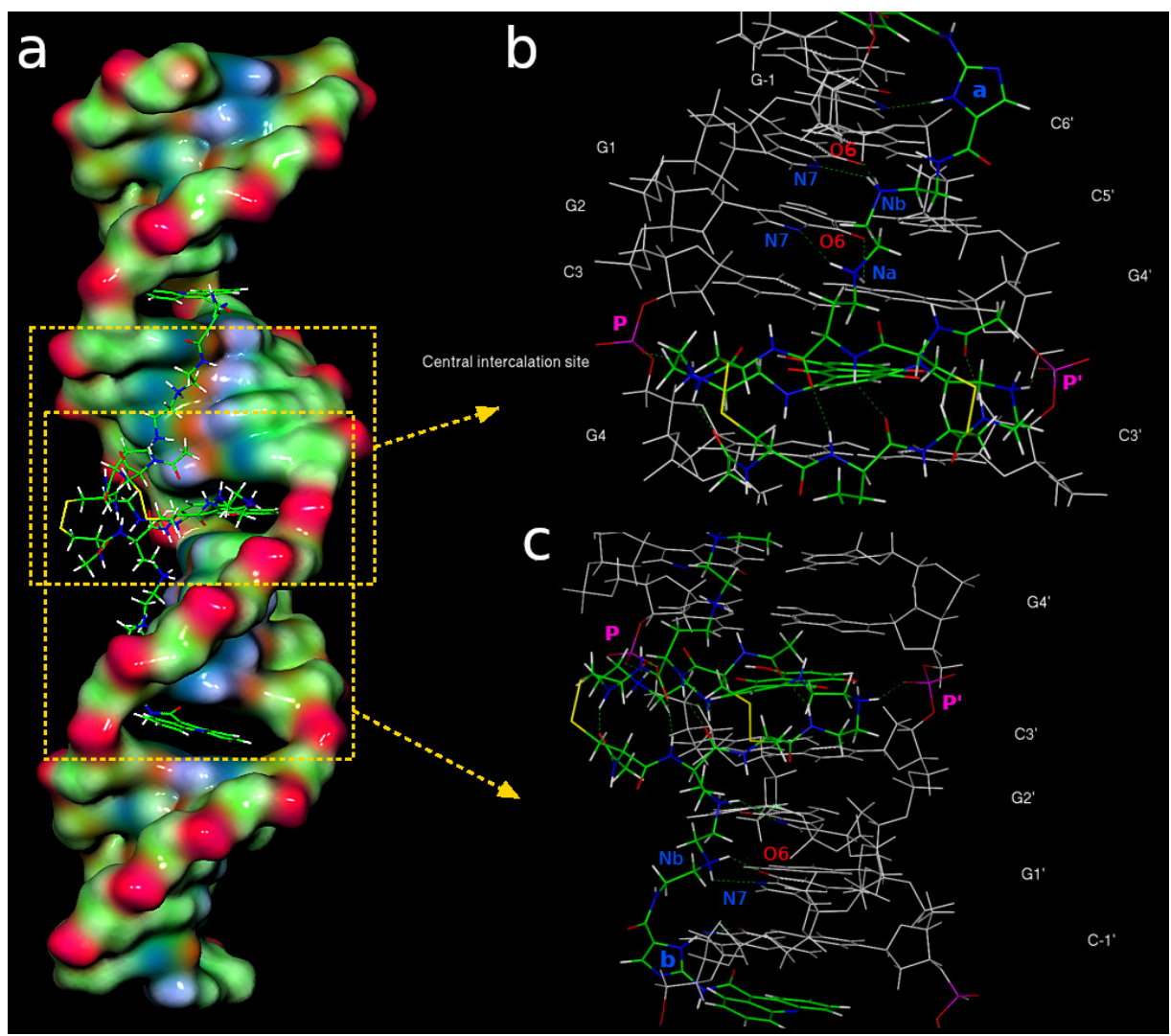

**Figure 6.** Compound **III** in complex with d(CGTA**C G**GGC G**CCC G**TACG)$_2$. **a)** overall view of the complex. Zoom on the upper **(b)** and lower **(c)** groove binding chains. Relevant atoms and nucleic bases are indicated.

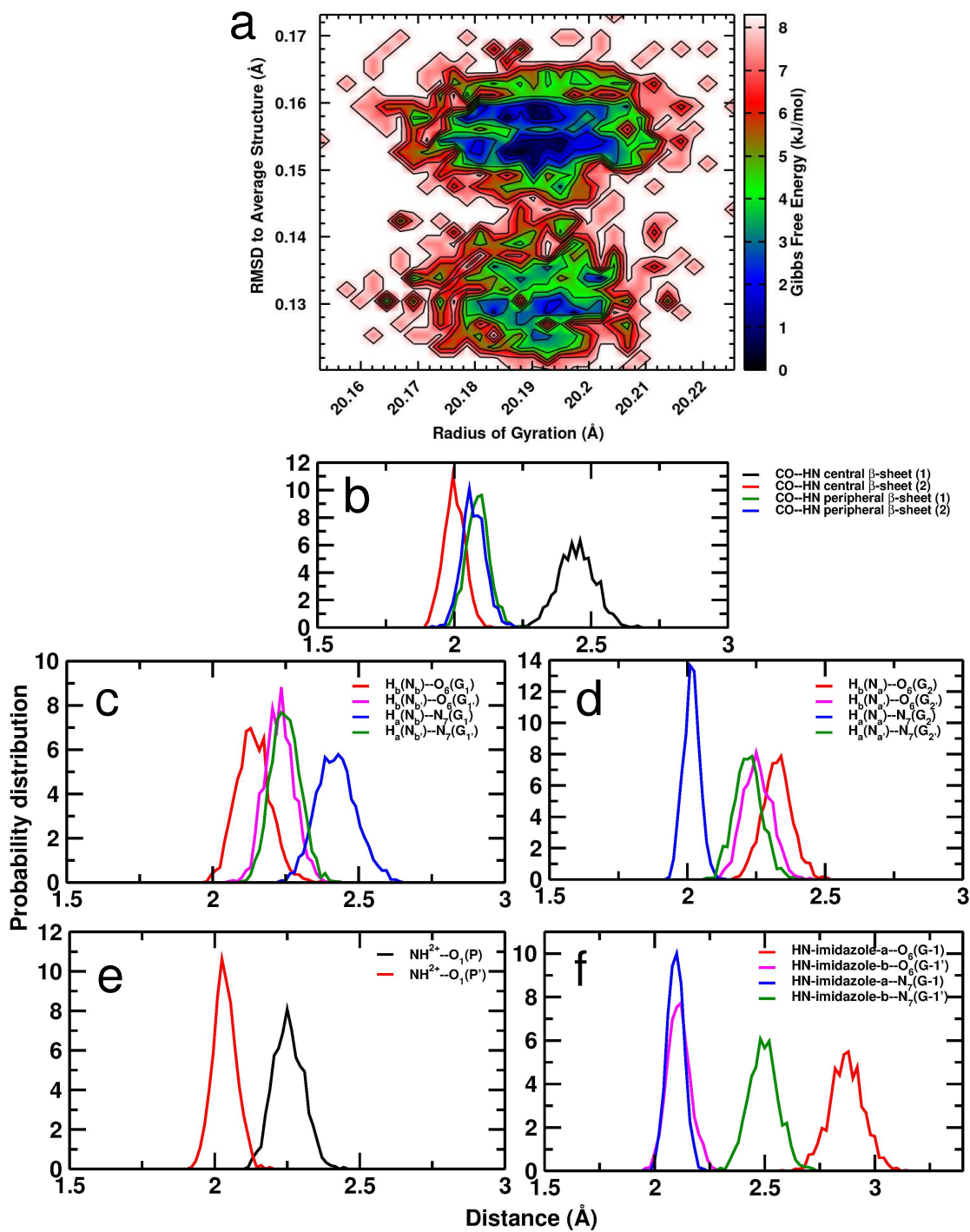

**Figure 7.** Structural investigation of **III** complexed with the target 18-mer oligonucleotide. **a)** Free energy landscape of the complex along the MD simulations; the zero energy is at 0 kcal/mol and corresponds to the lowest energy conformational state. **b-f)** Probability distribution of the following characteristic distances between **III** and the target: **b)** β−sheet stabilizing main-chain CO—HN distances around the dialkylammonium side-chain and in the N- and C-termini; **(c)** $O_6/N_7(G_1/G_{1'})$ with $H_a/H_b(N_b/N_{b'})$; **(d)** $O_6/N_7(G_2/G_{2'})$ with $H_a/H_b(H_a/N_{a'})$;

**(e)** ammonium N of the AMT carrier and $O_1$ of the central phosphate group facing it; **(f)** HN of the imidazole-amide connector with $N_7$ and $O_6$ of the 5'guanine of the 9-aminoacridine intercalation site, unprimed strand and primed strand. The corresponding time series are also reported in Supplementary Information Figure S3.

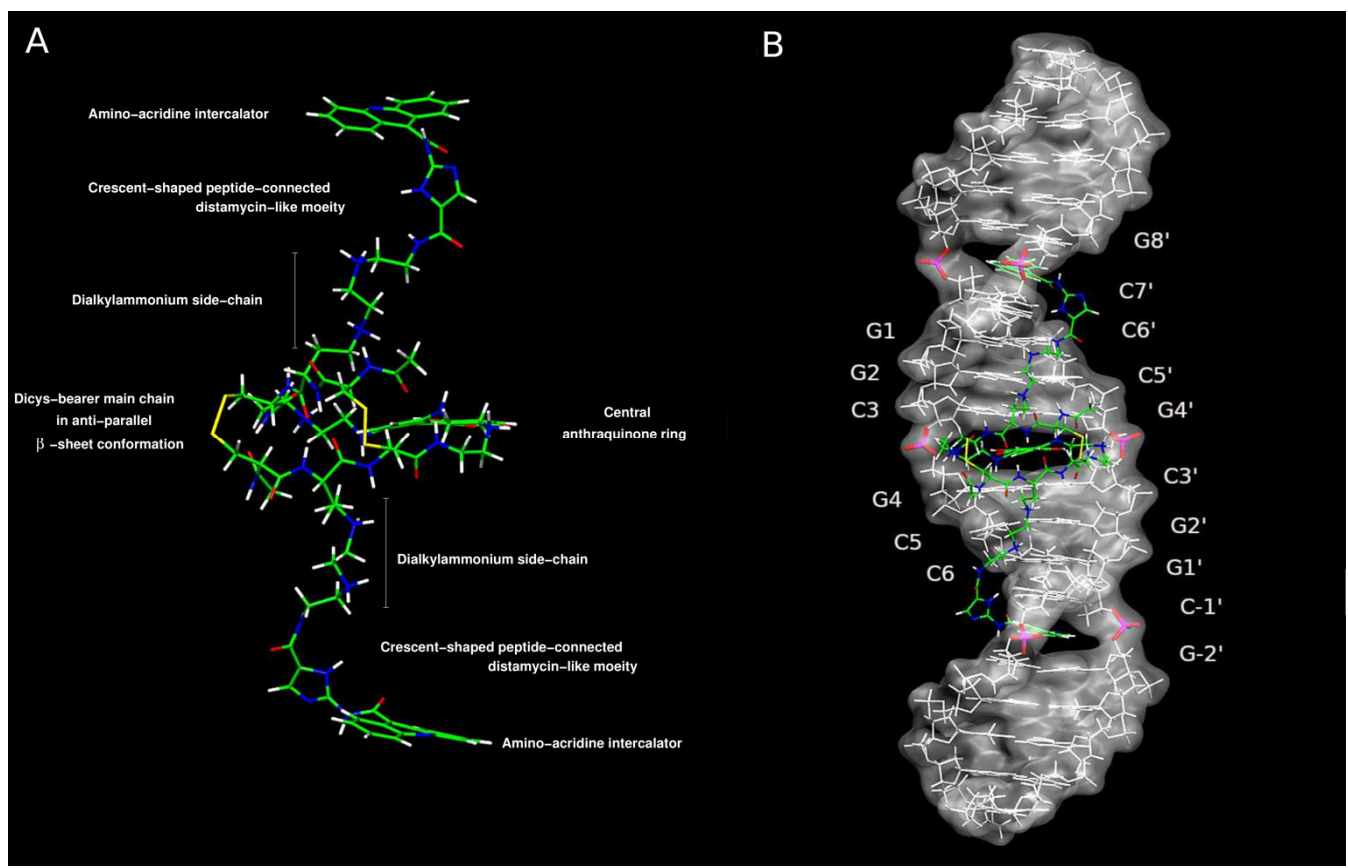

**Figure 8.** Figure representing the DNA-targeting interactions to tailor the trisintercalator compound **III**. **(a)** Chemical structure of compound **III** where the major chemical moieties are indicated. **(b)** Compound **III** in complex with its target d(CGTA**C GGGC G**CCC **G**TACG)$_2$. Phosphates of the 3 intercalation sites are shown in licorice and colored in magenta for the phosphate and in red for the oxygens.

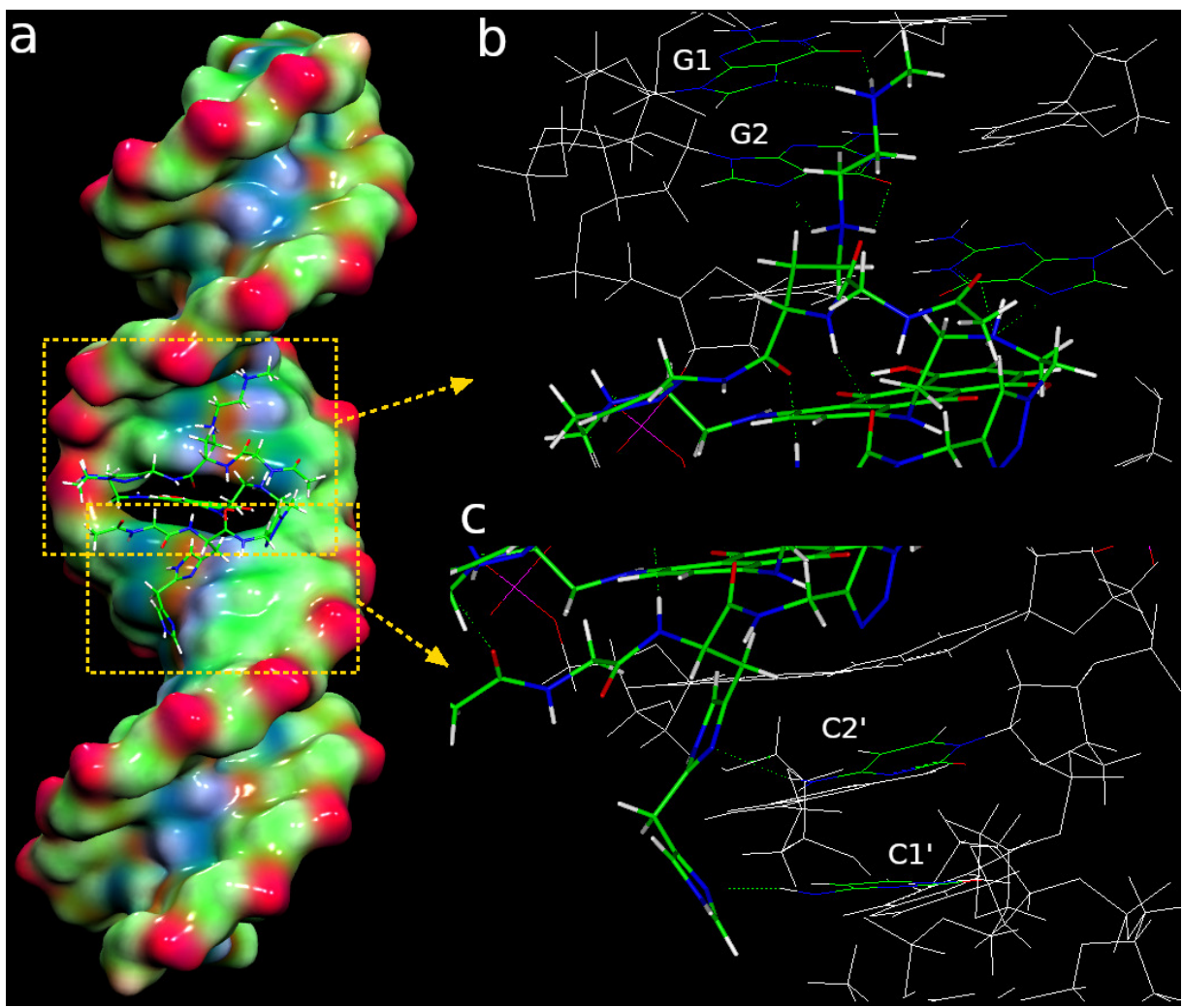

**Figures 9.** Compound **IV** in complex with the non-palindromic sequence d(GGC GGG).d(CCC GCC) in the target 18-mer oligonucleotide. **a)** overall view of the complex. Zoom on the upper **(b)** and lower **(c)** groove binding chains.

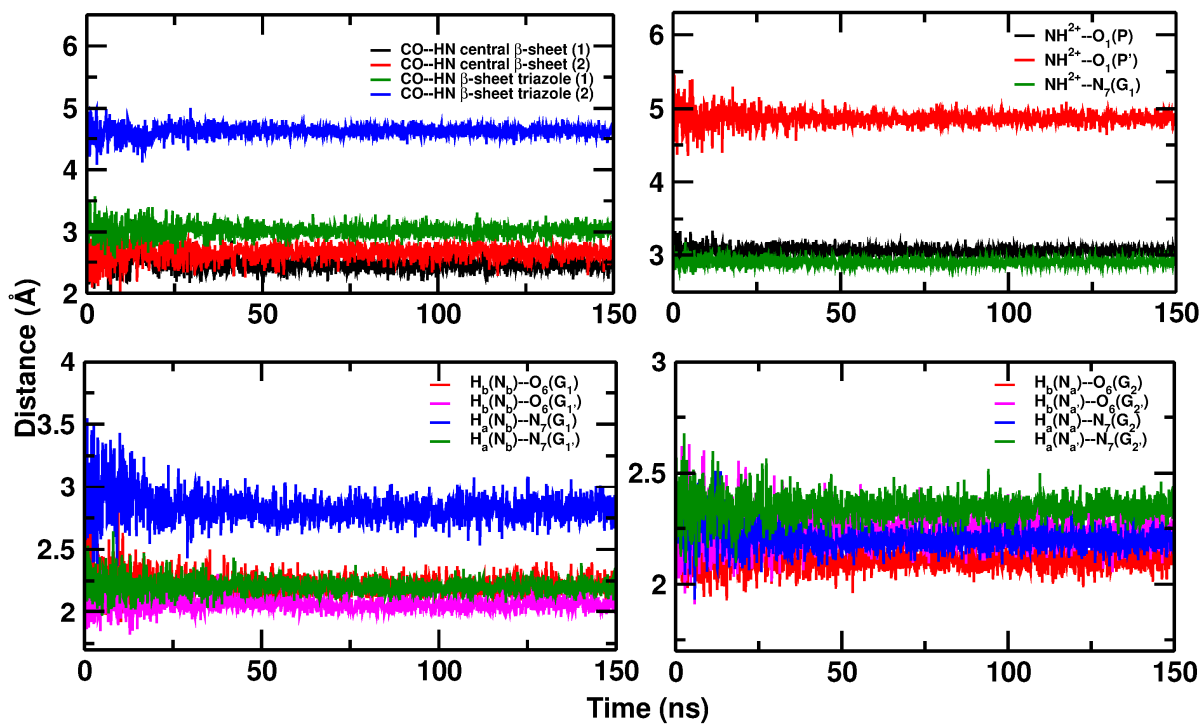

**Figure S1.** Time evolutions of characteristic distances between **I** and the target 18-mer oligonucleotide. **Upper left panel:** β−sheet stabilizing main-chain CO—HN distances around the dialkylammonium side-chain and CO-HC(triazole); **upper right panel:** ammonium N of the AMT carrier and $O_1$ of the central phosphate group facing it. The corresponding distance evolutions between the ammonium N on the primed site and $N_7$ of the 5' guanine of the intercalation site is also reported; **lower left panel:** $O_6/N_7(G_1/G_{1'})$ with $H_a/H_b(N_b)$; **lower right panel:** $O_6/N_7(G_2/G_{2'})$ with $H_a/H_b(N_a/N_{a'})$.

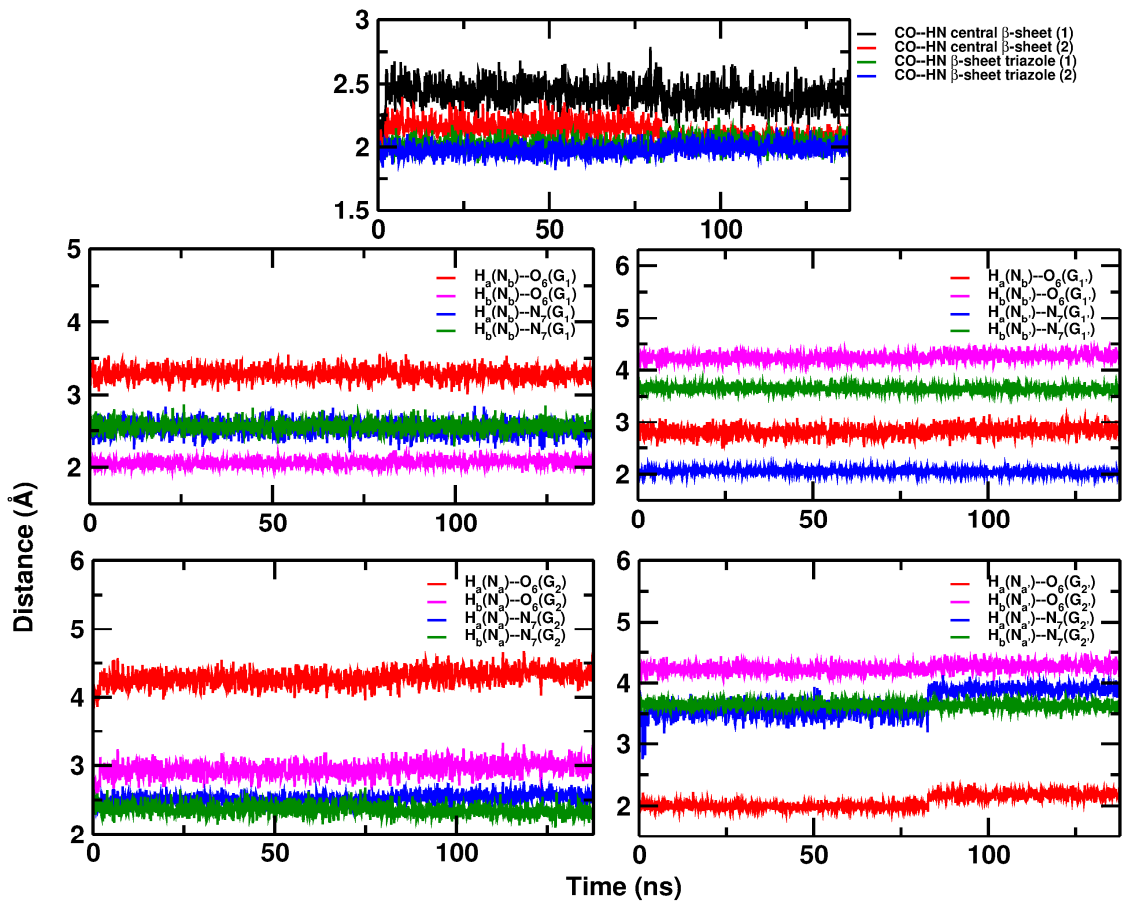

**Figure S2.** Time evolutions of characteristic distances between **II** and the target 18-mer oligonucleotide. **Top panel:** β−sheet stabilizing main-chain CO—HN distances around the dialkylammonium side-chain and in the N- and C-termini; **middle left panel:** $O_6/N_7(G_1)$ with $H_a/H_b$ ($N_b$); **middle right panel:** $O_6/N_7(G_{1'})$ with $H_a/H_b$ ($N_{b'}$); **lower left panel:** $O_6/N_7(G_2)$ with $H_a/H_b(N_a)$; **lower right panel:** $O_6/N_7(G_{2'})$ with $H_a/H_b(N_{a'})$.

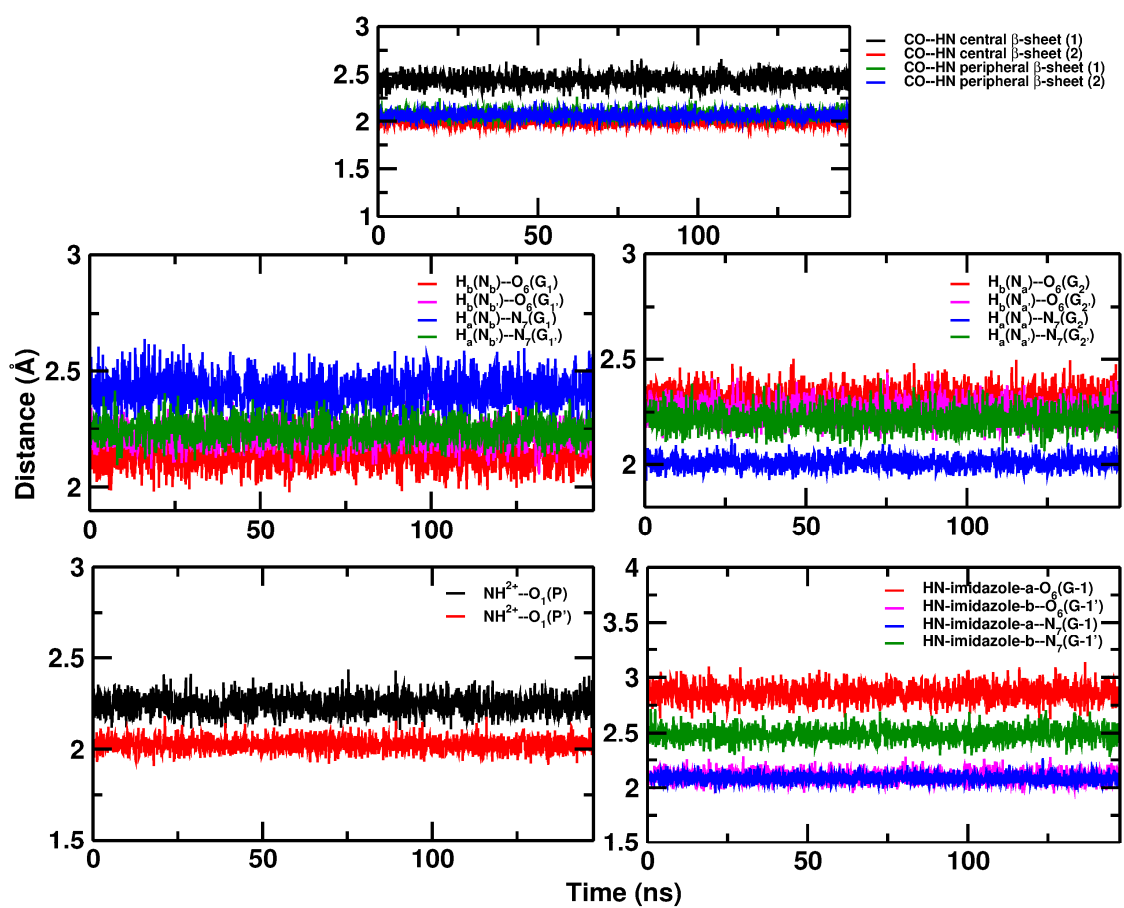

**Figures S3.** Time evolutions of characteristic distances between **III** and the target 18-mer oligonucleotide. **Top panel:** β−sheet stabilizing main-chain CO—HN distances around the dialkylammonium side-chain and in the N- and C-termini; **middle left panel:** $O_6/N_7(G_1/G_{1'})$ with $H_a/H_b(N_b/N_{b'})$; **middle right panel:** $O_6/N_7(G_2/G_{2'})$ with $H_a/H_b(H_a/N_{a'})$; **lower left panel:** ammonium N of the AMT carrier and $O_1$ of the central phosphate group facing it; **lower right panel:** HN of the imidazole-amide connector with $N_7$ and $O_6$ of the 5'guanine of the 9-aminoacridine intercalation site, unprimed strand and primed strand.